\documentclass[journal]{IEEEtran}
\input{psfig}

\usepackage{color,multicol}
\usepackage{amsmath}
\usepackage{amsfonts}
\usepackage{url}
\usepackage{amssymb}
\usepackage{fontenc}
\usepackage{times}
\usepackage{mathptmx}
\usepackage{graphicx}
\usepackage{subfig}
\usepackage{float}
\usepackage{xspace}
\usepackage{algorithmwh}

\newcommand{\mbf}[1]{\ensuremath{\mathbf{#1}}}
\newcommand{\mrm}[1]{\ensuremath{\mathrm{#1}}}

\begin{document}
\title{Efficient Low Dose X-ray CT Reconstruction through Sparsity-Based MAP Modeling}
\author{SayedMasoud Hashemi,~\IEEEmembership{Student Member, ~IEEE}, Soosan Beheshti,~\IEEEmembership{Senior Member, ~IEEE}, Patrick R. Gill, Narinder S. Paul, Richard S.C. Cobbold,~\IEEEmembership{Life Member, ~IEEE}
\thanks{Masoud Hashemi (Email: sayedmasoud.hashemiamroabadi@mail.utoronto.ca),
and Richard S.C Cobbold (cobbold@ecf.utoronto.ca) are with the Institute of Biomaterials and Biomedical Engineering (IBBME),
University of Toronto (UofT), Toronto, Canada. Soosan Beheshti (Email: soosan@ee.ryerson.ca) is with the Department of Electrical and Computer Engineering, Ryerson
University, Toronto, Canada. Patrick. R. Gill is with Rambus Inc., Sunnyvale, CA, USA.
Narinder S. Paul (Narinder.Paul@uhn.ca) is with University Health Network, joint Department of
Medical Imaging and the Institute of Biomaterials and Biomedical Engineering, University of Toronto, Toronto, Canada. }}
\maketitle

\begin{abstract}
Ultra low radiation dose in X-ray Computed Tomography (CT) is an important
clinical objective in order to minimize the risk of carcinogenesis.
Compressed Sensing (CS) enables significant reductions in radiation dose
to be achieved by producing diagnostic images from a limited number of CT projections.
However, the excessive computation time that conventional
CS-based CT reconstruction typically requires has limited clinical implementation.
 In this paper, we first demonstrate that a thorough
analysis of CT reconstruction through a Maximum a Posteriori objective
function results in a weighted compressive sensing problem. This analysis
enables us to formulate a low dose fan beam and helical cone beam CT
reconstruction. Subsequently, we provide an efficient solution to
the formulated CS problem based on a Fast Composite Splitting
Algorithm-Latent Expected Maximization (FCSA-LEM) algorithm.
In the proposed method we use pseudo polar Fourier transform as the measurement matrix
in order to decrease the computational complexity; and rebinning of the projections to parallel rays in order to extend its
application to fan beam and helical cone beam scans. The weight involved
in the proposed weighted CS model, denoted by Error Adaptation Weight
(EAW), is calculated based on the statistical characteristics of CT
reconstruction and is a
function of Poisson measurement noise and rebinning interpolation error.
Simulation results show that low computational complexity of the proposed method made
the fast recovery of the CT images possible and using EAW  reduces the reconstruction error
by one order of magnitude. Recovery of a high quality 512$\times$ 512 image was achieved in less than 20 sec
on a desktop computer without numerical optimizations.
\end{abstract}
\begin{keywords}
Computed Tomography, Direct Fourier Reconstruction, Pseudo-Polar Fourier Transform, Compressed Sensing, Statistical
Iterative CT reconstruction
\end{keywords}

\section{Introduction}
The clinical use of Computed Tomography (CT) has dramatically increased over
the last two decades. This is primarily due to its unsurpassed speed and the
fine details that can be obtained in cross-sectional views of soft tissues and organs. Compared to conventional radiography, CT results in a relatively
large radiation dose to patients. Studies over the past decade have shown
that the higher radiation dose is of serious long-term concern in its potential for increasing the risk of developing cancer \cite{Brenner2007,Pearce}. As a result, low dose CT  imaging that maintains
the resolution and achieves good contrast to noise ratio has been the goal of many CT developments over the past decade.\\
Low dose CT images reconstructed with conventional Filtered Back Projection (FBP), which directly calculates the image in a single reconstruction step,
suffer from low contrast to noise ratios.
A reduced radiation dose decreases either the number of emitted photons or their energy. This increases the amount of photon noise in CT images and degrades the image quality.
Several methods have been proposed for lowering the relative amount of noise in a low dose CT scan.
These methods can be categorized into the following three different approaches:
1) improving scan protocols \cite{ECGgating},
2)  adding denoising algorithms \cite{denoise1,denoise2},
and 3) investigating new reconstruction methods \cite{SIR,SART,Sidky2006,Sidky2009,PICCS2008,PICCS2012}.
The first approach is hardware based.
New iterative reconstruction approaches are proposed by
combining the goals of the second and third approaches.
These methods aim to improve the reconstruction quality and
to decrease  image artifacts.
Iterative reconstruction methods can be categorized into two groups:
Algebraic Reconstruction Techniques (ART) \cite{SART,randkaczmarz1,randkaczmarz2} and Statistical Iterative Reconstruction (SIR) \cite{SIR}.
While SIR methods are more successful in noisy (low dose) reconstructions, ART based methods
have advantages in dealing with incomplete data. However, compared to FBP
both methods are computationally expensive enough to hinder their widespread clinical adoption. \\
Iterative reconstruction methods have progressed with the introduction of Compressed Sensing (CS).
CS is a relatively recent innovation in signal processing that allows recovery of images from fewer
projections than that required by the Nyquist sampling  theorem \cite{Candes2006,Donoho2006}.
The overall X-ray exposure in CT scanners is the product of the X-ray exposure at each projection view and the number of projection views.
While conventional iterative CT reconstruction methods focus on reducing the amount of X-ray exposure in the projections,
CS permits reconstructions from fewer X-ray projection views.
Such methods are capable of reconstructing high quality images from  approximately one tenth of the number of views needed in FBP \cite{Sidky2006}, permitting a much lower dose scanning protocol than than that needed in conventional reconstruction methods.
However, CS-based reconstruction/tomography algorithms suffer from two drawbacks: they are prohibitively computationally
intensive for clinical use \cite{PICCS2008,PICCS2012}, and they have not incorporated CT statistics and geometries in
problem formulation \cite{RecPF,Fahimian2010,MRIFCSA,TwIST}.
As a result, it seems unlikely that these methods could be used directly for the clinical CT systems. \\
CS prescribes solving optimization problems such as those given by:
\begin{eqnarray}
\hat{\mbf{x}}&=&\underset{\mrm{x}}{\operatorname{argmin}}\,\frac{1}{2}\|\mbf{y}-\mbf{Ax}\|^2_2+\lambda_{R} \mbf{TV}(f) \label{eqn:TVmin} \\
\hat{\mbf{x}}&=&\underset{\mrm{x}}{\operatorname{argmin}}\,\frac{1}{2}\|\mbf{y}-\mbf{Ax}\|^2_2+\lambda_{R} \|W^T\mbf{x}\|_1 \label{eqn:L1min}
\end{eqnarray}
where $\lambda_R$ acts as a regularization parameter specifying a trade-off between the image prior model and
the fidelity to observations, $\mbf{A}$ is the measurement matrix, $\mbf{x}$ is the column vector representation of
the desired image ($f$), $\mbf{y}$ is the measured data, $W^T$ is a sparsifying transform, $\|x\|_2^2 = \sum_{i} |x_i|^2$, $\|x\|_1 = \sum_{i} |x_i|$, and
$\mbf{TV}(f)=\sum_i\sqrt{(\nabla_xf)^2_i+(\nabla_yf)^2_i}$ where $\nabla_x$ and
$\nabla_y$ are the first derivatives in direction $x$ and $y$ accordingly. \\
The main challenge in solving this optimization problem within a reasonable amount of time  is due to the size of the measurement matrix \mbf{A}.
Currently, in most available CS-based reconstruction methods used for modern CT geometries the measurement matrix \mbf{A} is a Radon sampling matrix which models the rays going through the patient.
For example, to reconstruct a $512 \times 512$ pixel image from 900 sensors and 1200 projection angles, $\mbf{A}$ would be a $1080000 \times 262144$ matrix.  As typical iterations
each usually require two multiplications by $\mbf{A}$ and $\mbf{A^T}$, it takes several hours of computation on typical desktop computers to reconstruct a $512\times512$
image with such methods \cite{PICCS2008,PICCS2012}. \\
To reduce the computational complexity of the
image reconstruction, Fourier based reconstruction methods have been proposed \cite{DFR1,DFR2,DFR3}.
The Central Slice Theorem (CST) or Direct Fourier Reconstruction (DFR)
relates the 1D Fourier transform of the projections
to the 2D Fourier transform of the image.
DFR reconstructions comprise:
1) interpolation of  polar data to a Cartesian grid and
2) calculation of the inverse FFT on that grid to reconstruct the CT image.
Moreover, to achieve an acceptable reconstruction quality, the interpolation step needs
oversampling, which requires additional radiation exposure. \\
To address the interpolation problem in DFR based methods,
Equally Sloped Tomography (EST) has been proposed \cite{Mao2010,Miao2005,Lee2008}.
EST is an iterative method using the Pseudo Polar Fourier transform (PPFT) \cite{Averbuch2006}.
The PPFT has three important properties which makes it a good alternative to conventional DFR methods:
1) it is closer to polar (equiangular line) grids compared to Cartesian grids,
2) it can be computed with a fast algorithm \cite{Averbuch2006}, and 3) unlike interpolating the polar data on a Cartesian grid in regular DFR methods it has an analytical conjugate function. \\
Note that all the above mentioned methods such as CST, DFR, and EST
assume parallel beam geometry and do not take account of the fan and cone beam geometries used in most current CT systems.
Consequently, in order to use them
the projected rays should first be transformed to parallel beams.
This step, called rebinning, includes interpolation that induces additional error to the reconstructed image.
This problem has received slight attention, although it has been addressed in the following two papers.
An EST based method was proposed to reconstruct fan beam and helical cone beam images in \cite{fahimian2013}.
In this method, to overcome the rebinning interpolation problem, at each iteration a non-local total variation minimization smoothing step is used.
An $\ell_2$-TV optimization scheme was used to reconstruct the CT images from fan beam projections in \cite{hashemi2013}.
To compensate the interpolation error, a confidence matrix is added to
the CS scheme, which controls the propagation of the error in the iterations.\\
It should be noted that while
the geometry can be incorporated into CS-based reconstruction by rebinning,
the statistics of the noise that typically occurs in CT data has not been utilized in formulating the problem, such as those given by (\ref{eqn:TVmin}) and (\ref{eqn:L1min}).
A modified CS formulation, called reweighted $\ell_1$ minimization, was proposed in \cite{weightedCS3} where
it was discussed that using appropriate weights the quality of the recovered signal can be improved.
Using the weights introduced in this modified formulation, some statistical priors can be added to the model. For example, two weighting methods were proposed in \cite{weightedCS1} and \cite{weightedCS2} for
recovery of the signals with a partial known support and with \emph{a priori} information about the probability of each entry of the signal being non-zero. \\
In this paper, we rigourously explore the statistical characteristics of CT image reconstruction to model it based on Maximum a Posteriori (MAP) formulation.
It is shown that the MAP formulation is transformed into a weighted CS problem. The weights are
direct consequences of the geometry and statistics of CT itself. The resulted weight, denoted by Error Adaptation Weight (EAW),
is a function of Poisson measurement noise and the interpolation error caused by rebinning.
The first part of this paper leads to a proposed weighted CS problem, which is solved by the method proposed in the second part.
To provide an efficient solution,
we first break the optimization problem into two simpler $\ell_2-\ell_1$ and $\ell_2$-TV problems by using Fast Composite Splitting Algorithm (FCSA) \cite{FCSA}.
Next we solve each optimization problem with a latent Expectation-Maximization (LEM) method. The overall solution, denoted by FCSA-LEM,
is able to reconstruct high quality images from fewer projections and consequently lower-dose CT scans while using substantially less computation load than
 conventional methods.\\
The paper is organized as follows. In section \ref{sec:form} the CT reconstruction problem is formulated and a MAP model is introduced for CT images.
In section \ref{sec:Prep} we discuss the procedure of how to transform the regularized CT inverse problem into one that can be solved quickly and with few interpolation artifacts.
The proposed image reconstruction algorithm based on the EM estimator is provided in section \ref{sec:EM-MAP}, and
section \ref{sec:results} contains the simulation results.
% ======================================================================================================================
%                                                       MAP
% ======================================================================================================================
\section{Problem Formulation} \label{sec:form}
In this section we describe a CT reconstruction method based on the optimization of the
maximum a posteriori of the projection data and given priors.
We use two classes of prior: sparsity of the wavelet coefficients and piecewise linearity of the images,
and will show that the proposed MAP model is similar to weighted CS model.
The key innovation of our method is the introduction of  weights applied to the data, which depend on the magnitude of the noise sources from both measurement and interpolation.
\subsection{Maximum a Posteriori Model (MAP) of CT} \label{sec:MAP}
X-ray projections of the parallel beam CT can be expressed as the Radon transform of the object. The Radon transform is defined as \cite{FBP}:
\begin{equation}\label{eqn:radon}
    g(l,\varphi)=\mathcal{R}(f)=\int_{-\infty}^{\infty}\int_{-\infty}^{\infty}f(u,v)\delta(u\cos\varphi+v\sin\varphi-l)dudv
\end{equation}
which is the integral along a ray at angle $\varphi$ and at the distance $l$ from the origin, $\delta(u,v)$ is Dirac delta function, and $f(u,v)$ is the
image attenuation at $(u,v)$.\\
However, this is not what the scanners directly measure.
Detectors of the scanner measure the number of photons which hit the detector in different angles, $\lambda(l,\varphi)$, which
is usually modeled by Poisson distribution with expected value of $\overline{\lambda}(l,\varphi)$ \cite{CTPrinc}. The relation between the projections, $g(l,\varphi)$, and $\overline{\lambda}(l,\varphi)$ is given by:
\begin{equation}\label{eqn:projandlmbda}
\overline{\lambda}(l,\varphi) = \lambda_T \exp(-g(l,\varphi))
\end{equation}
in which $\lambda_T$ is the number of radiated photons from the X-ray source. This leads to:
\begin{equation}\label{eqn:projandlmbda2}
g(l,\varphi) = -\log(\frac{\overline{\lambda}(l,\varphi)}{\lambda_T})
\end{equation}
$\lambda(l,\varphi)$ is usually corrupted with two kinds of noise: electrical noise of the detectors (with variance of $\sigma_n^2$) and the photon counting noise
(observed counts are drawn from a Poisson distribution of mean $\bar{\lambda}$).
If we consider the discrete formulation in which $\mbf{y}$ denotes the vectorized $g(l,\varphi)$, $\mbf{x}$ denotes the
vectorized $f(x,y)$, and \mbf{A} is the projection matrix,
using the second order Taylor series expansion of the Poisson distribution and log likelihood of the measurements, we have \cite{MAPmodel1,MAPmodel2}:
\begin{equation}\label{eqn:taylor}
\log p(\mbf{y}|\mbf{x}) \approx -\frac{1}{2} (\mbf{y}-\mbf{A}\mbf{x})^T D (\mbf{y}-\mbf{A}\mbf{x}) + O(\mbf{y}^3)
\end{equation}
in which $O(\mbf{y}^3)$ is a function which depends upon measured data only and $D$ is a diagonal matrix.  The $i^{th}$ diagonal element of
$D$ is denoted by $d_i$.
Ignoring $O(\mbf{y}^3)$, (\ref{eqn:taylor}) describes a simplified CT model:
\begin{equation}\label{eqn:GausModel}
  \mbf{y} = \mbf{A}\mbf{x}+\mbf{n}
\end{equation}
in which \mbf{n} is a Gaussian distributed noise with a covariance matrix $D^{-1}$ and
$d_i$ is proportional to detector counts which are the maximum likelihood of the inverse of the variance of the projection measurements,
$1/\sigma_y^2$. From (\ref{eqn:projandlmbda2}) the relation for the $i^{th}$ measured projection $\mbf{y}_i$ is:
\begin{eqnarray}\label{eqn:di}
\mbf{y}_i = \log (\frac{\lambda_T}{\lambda_i}) = \log (\frac{\lambda_T}{\overline{\lambda}_i}) + \log (\frac {\overline{\lambda}_i}{\lambda_i}) \nonumber \\
\approx \overline{\mbf{y}}_i + (1-\frac {\lambda_i}{\overline{\lambda}_i})
\end{eqnarray}
in which $\overline{\mbf{y}}_i$ is noiseless and $\lambda_i$ follows the Poisson distribution with $\sigma_{\lambda}^2 = \overline{\lambda}_i$.
As a result the variance of projection data can be estimated from:
\begin{equation}\label{eqn:yistd}
\sigma_{\mbf{y}_i}^2 \approx (\sigma_{\lambda}^2 + \sigma_n^2)(\frac{1}{\overline{\lambda}_i})^2
\end{equation}
Using $\lambda_i$ as an unbiased estimation of $\overline{\lambda}_i$ the diagonal elements of $D$ can be expressed as:
\begin{equation}\label{eqn:di}
%D = \begin{bmatrix} d_1 & \cdots & 0 \\ \vdots & \ddots & \vdots \\ 0 & \cdots & d_{(k  n_\varphi) \times n^2} \end{bmatrix}, d_i = \frac{1}{\sigma_{\mbf{y}_i}^2}= \frac{\lambda_i^2}{\sigma_n^2+\lambda_i}
d_i = \frac{1}{\sigma_{\mbf{y}_i}^2}= \frac{\lambda_i^2}{\sigma_n^2+\lambda_i}
\end{equation}
Therefore, the MAP estimator can be used to reconstruct the image from the projections, which uses the following equation:
\begin{equation}\label{eqn:MAP}
\hat{\mbf{x}} = \underset{\mrm{x}}{\operatorname{argmax}}\, {\log p(\mbf{y}|\mbf{x}) + \log p(\mbf{x})}
\end{equation}
Here $h(x) = \log p(\mbf{x})$ acts as a penalty function, which will be used later in the paper.
It has been shown in many studies \cite{hashemi2012,GGD} that the wavelet transform of the natural and medical images, $\theta = W^T\mbf{x}$,
can be modeled by Generalized Gaussian Distribution (GGD):
\begin{equation}\label{eqn:GGD}
  p(\theta) = p(W^T\mbf{x}) = K(s,q) \cdot \exp (-|\frac{\theta}{s}|^q)
\end{equation}
where $W^T$ is a sparsifying transform such as the wavelet transform and $W$ is its inverse, $s$, $q$ are the parameters of the GGD and $K(s,q)$ is the normalization parameter. When $q=1$, the GGD is equivalent to
Laplacian distribution and when $q=2$ it describes a Gaussian distribution. Using (\ref{eqn:GGD}), (\ref{eqn:taylor}) and (\ref{eqn:MAP})
we have the following MAP model for CT images:
\begin{equation} \label{eqn:modelL1}
 \hat{\mbf{x}} = \underset{\mrm{x}}{\operatorname{argmin}}\, \frac{1}{2} (\mbf{y}-\mbf{A}\mbf{x})^T D (\mbf{y}-\mbf{A}\mbf{x}) + \lambda_R \|W^T\mbf{x}\|_q
\end{equation}
Typically, $q$  is chosen to be $0 < q \leq 1$, $\theta$ is a sparse representation of the image $\mbf{x}=W\theta$, and $\|x\|_q = \sum_i |x_i|^q$.

Another prior on $p(x)$ is the piecewise linearity of the images. A $p$-variation distribution is proposed to describe
piecewise constant functions \cite{TVprior}. If $x_n(t) = \sum_{j=1}^n x_j^n \psi_j^n(t)$ is a piecewise function spanned by $\psi_j^n(t)$, the
roof-top basis, the following class of probability distribution can be used to describe it:
\begin{eqnarray}\label{eqn:pvariation}
  p(x_1^n,...,x_n^n) = c_{q,n} \exp (-\frac{a_n}{(n+1)^{1-q}} \sum_{j=1}^{n+1}|x_j^n-x_{j-1}^n|^q)
\end{eqnarray}
where $a_n > 0$, $x_0^n = x_{n+1}^n = 0$, $c_{q,n}$ is normalizing factor, and $[ x_1^n,...,x_n^n ]^T$ is a $\mathbb{R}^n$-valued random vector.
When $q=1$, this yields the total variation norm. Using (\ref{eqn:pvariation}) with $q=1$, (\ref{eqn:taylor}) and (\ref{eqn:MAP})
become the following MAP model for CT images:
\begin{equation} \label{eqn:modelTV}
 \hat{\mbf{x}} = \underset{\mrm{x}}{\operatorname{argmin}}\, \frac{1}{2} (\mbf{y}-\mbf{A}\mbf{x})^T D (\mbf{y}-\mbf{A}\mbf{x}) + \lambda_R \mbf{TV}(f)
\end{equation}
 As can be seen, (\ref{eqn:modelTV}) and (\ref{eqn:modelL1}) are generalized forms of the CS models given by (\ref{eqn:TVmin})
and (\ref{eqn:L1min}), respectively. Consequently, applying CS for CT reconstruction is equivalent to a MAP estimation of the CT images.
%======================================================================================================================
%                                                   Pre-processing
% ======================================================================================================================
\section{Generalized CS Model for Fan Beam and Helical Cone Beam Geometries} \label{sec:Prep}
\begin{figure}[htb]
\begin{minipage}[b]{.48\linewidth}
  \centering
  \centerline{\includegraphics[width=4.0cm]{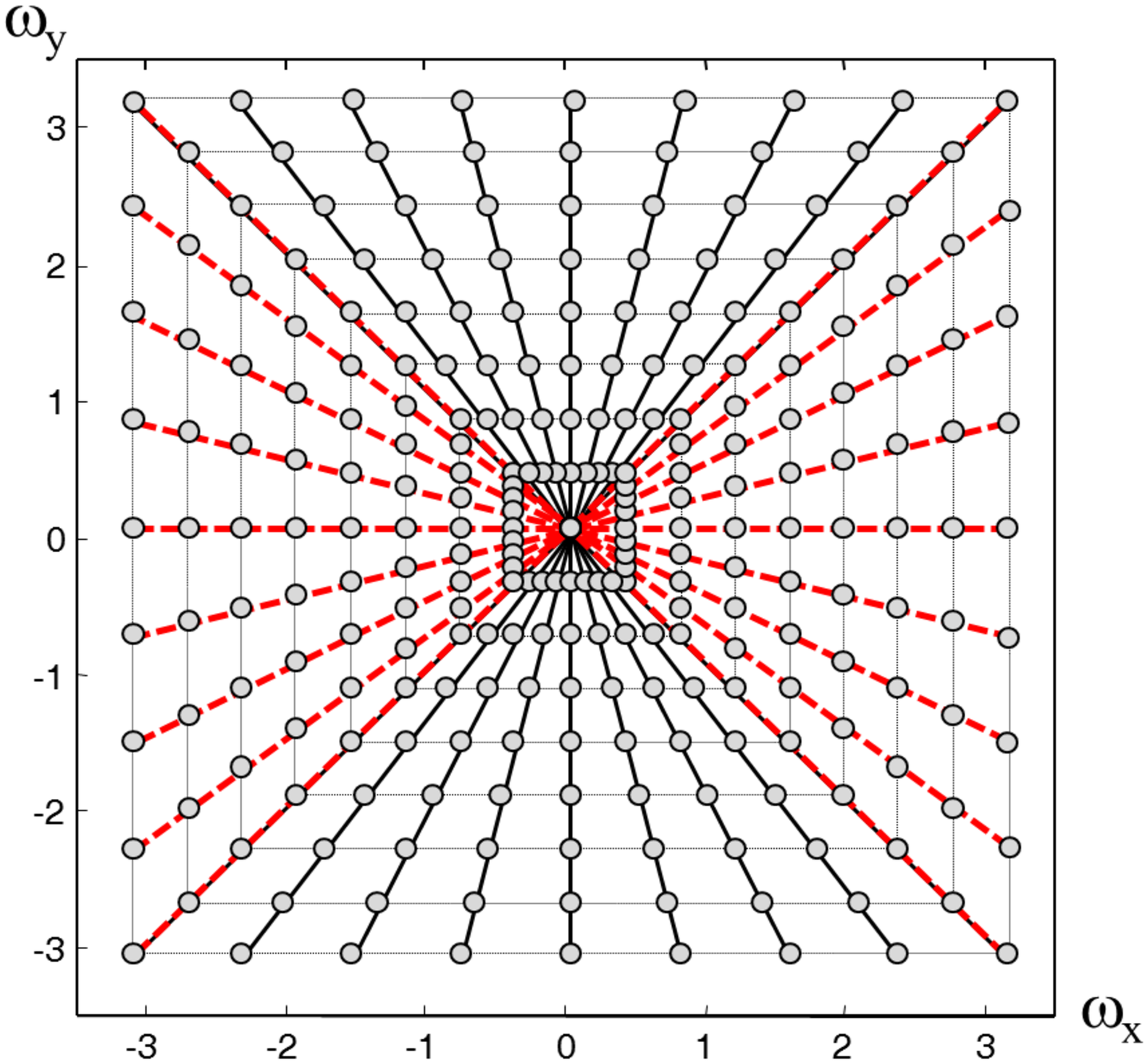}}
  \centerline{(A)}\medskip
\end{minipage}
\hfill
\begin{minipage}[b]{0.48\linewidth}
  \centering
  \centerline{\includegraphics[width=4.0cm]{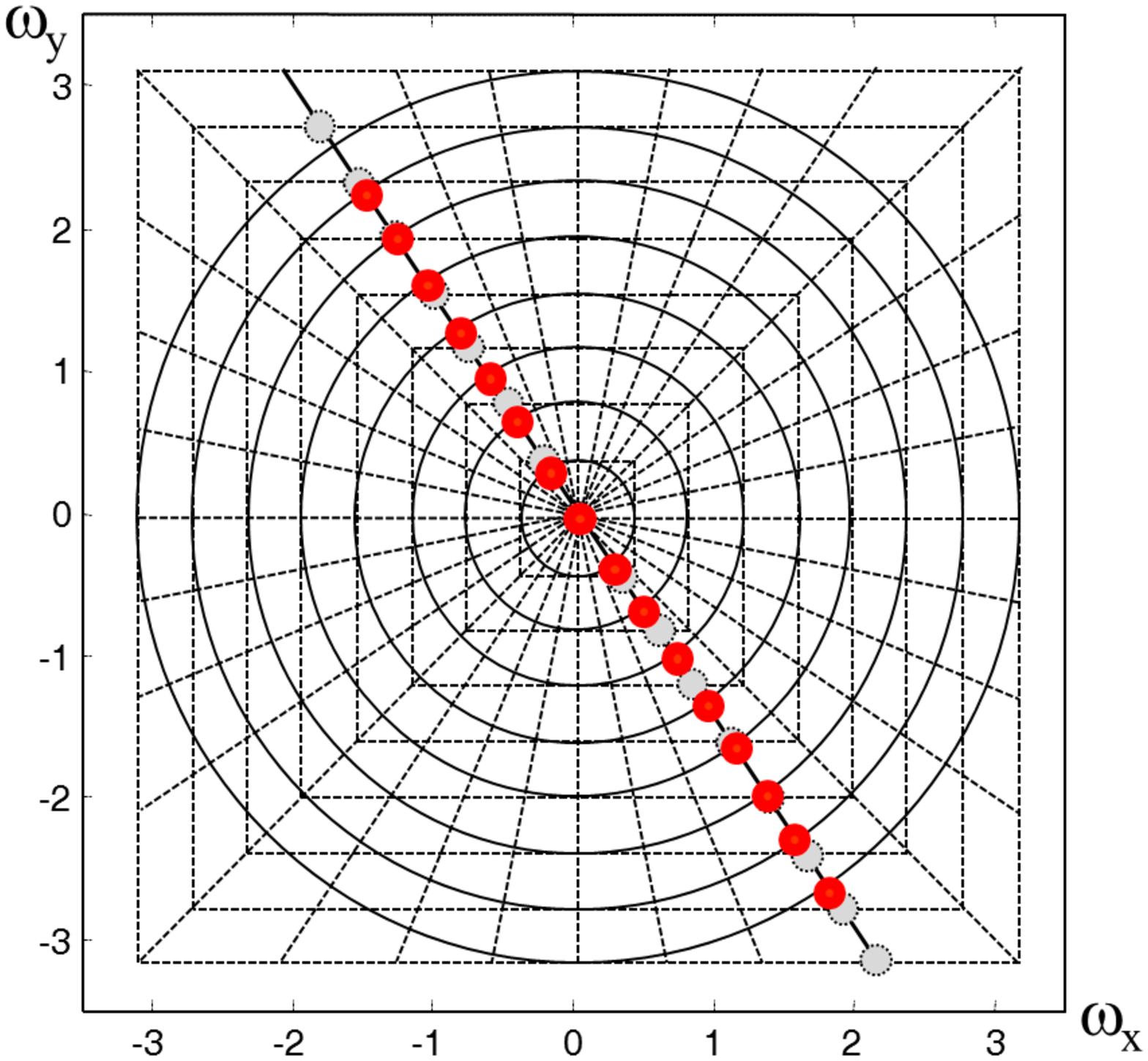}}
  \centerline{(B)}\medskip
\end{minipage}
\caption{(A) Pseudo-Polar Grids: red lines are Basically Horizontal (BH) and black lines are Basically Vertical (BV). (B) Polar Grids (red dots) on Pseudo-Polar Grids (gray dots).}
\label{fig:BHBV}
\end{figure}

To reduce the computational complexity of the CT reconstruction, the pseudo polar Fourier transform is used as the measurement matrix, \mbf{A}.
Since pseudo polar grids are placed on  equally sloped radial lines,
the projection rays should be measured or interpolated on the equally sloped radial lines, as shown in Figure \ref{fig:BHBV}:
%\begin{eqnarray}\label{eqn:ppftangles}
%  \varphi_{BH} &=& \tan^{-1}\frac{2m}{N}, -\frac{N}{2} \leq m <\frac{N}{2}  \nonumber \\
%  \varphi_{BV} &=& \tan^{-1}\frac{2m}{N}+\frac{\pi}{2},-\frac{N}{2} \leq m <\frac{N}{2}  \nonumber \\
%  \varphi &=& \varphi_{BH} \bigcup \varphi_{BV}
%\end{eqnarray}
\begin{eqnarray}\label{eqn:ppftangles}
  \varphi_{BH} &=& \tan^{-1}2m/N, -N/2 \leq m < N/2  \nonumber \\
  \varphi_{BV} &=& \tan^{-1}2m/N+\pi/2,-N/2 \leq m <N/2  \nonumber \\
  \varphi &=& \varphi_{BH} \bigcup \varphi_{BV}
\end{eqnarray}

\begin{figure}[htb]
\begin{minipage}[b]{.98\linewidth}
  \centering
  \centerline{\includegraphics[width=7cm]{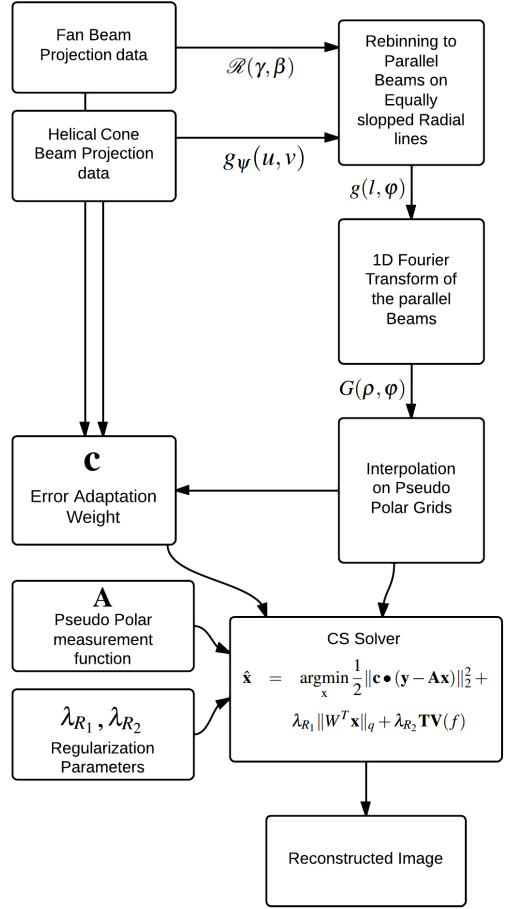}}
\end{minipage}
\caption{Flowchart of the reconstruction method.}
\label{fig:FanbeamFlowChart}
\end{figure}

Although CT scanners typically collect data along equally spaced angles, they have the flexibility to collect data along the angles of a pseudo polar grid instead.
 Then, the equally distant measured data should be interpolated to
 the pseudo polar girds, as shown in Figure \ref{fig:BHBV}-B. The resulting interpolation error can be limited by oversampling
the Fourier data by zero-padding the projections on the equally sloped radial lines.
Fan beam and helical geometries need extra interpolations to estimate the measured projections on the parallel equally sloped radial lines first, a process called rebinning.
At each interpolation step, the interpolation error is tracked to be included in the EAW.
The calculated weights and the prepared data are fed into the FCSA-LEM solver.
\\
Consequently, the proposed CT reconstruction method, shown in Figure \ref{fig:FanbeamFlowChart}, can be summarized by the following two major stages:
\begin{itemize}
  \item Data preparation and rebinning: fan or helical projections are mapped to  parallel equally sloped radial lines used
  in the pseudo polar Fourier transform. The output  \mbf{y} of this stage is the 1D Fourier transform of the calculated parallel rays.
  \item Image Reconstruction: the CT image is reconstructed using the proposed FCSA-LEM method. The measurement matrix is the fast pseudo polar Fourier transform function and the
  input data is \mbf{y} from the first stage.
\end{itemize}
%Data preparation and rebinning stage is discussed in this section.
%======================================================================================================================
%                                                       PPFT
% ======================================================================================================================
\subsection{Complexity Reduction through the use of the Pseudo Polar Fourier Transform (PPFT)}
Pseudo polar grids contain two types of samples: basically horizontal (BH) and basically vertical (BV), as seen
in Figure \ref{fig:BHBV}.  BV and BH lines are described by:
\begin{flalign}
  BV = \{ & \omega_y = \frac{\pi l}{N} \text{ for } -N \leq l <N,            \nonumber & \\
       &    \omega_x = \omega_y.\frac{2m}{N} \text{ for } -N/2 \leq m <N/2 \}    \nonumber & \\
  BH = \{ & \omega_x = \frac{\pi l}{N} \text{ for } -N \leq l <N,            \nonumber & \\
       &    \omega_y = \omega_x.\frac{2m}{N} \text{ for } -N/2 < m \leq N/2 \}             &
\end{flalign}
The Fourier transform on the BV grids can be found from:
\begin{equation}\label{eqn:BVFT}
  F(\omega_x,\omega_y) = F[m,l] = \sum_{i_1=0}^{N-1} \hat{f_1}[i_1,l] \exp({-\frac{i2\pi i_1 m}{N}.\frac{l}{N}})
\end{equation}
where $\hat{f_1}[i_1,l]=\sum_{i_2=0}^{2N-1}f_Z[i_1,i_2] (-1)^{i_2} \exp(-\frac{i2\pi i_2 l}{2N})$ is the 1D Fourier transform of the zero-padded columns of the image ($f_Z$).
In fact, the same equation can be written for BH by applying the same equation on rows rather than columns. As a result,
(\ref{eqn:BVFT}) can be interpreted as the fractional Fourier transform of the 1D Fourier transform of the zero-padded columns of the image weighted by $(-1)^{i_2}$.
To reconstruct an $N \times N$ image from its PFFT coefficients, $4N^2$ samples are needed.
A fast algorithm is proposed in \cite{Averbuch2006} to calculate the PPFT and its conjugate; it is
used in our proposed algorithm as \mbf{A} and $\mbf{A}^T$, respectively.
% ======================================================================================================================
%                                                       rebinning fan beam
% ======================================================================================================================
\subsection{Rebinning Process}\label{sec:rebinning}
\begin{figure}[htb]
\begin{minipage}[b]{.45\linewidth}
  \centering
  \centerline{\includegraphics[width=4.cm]{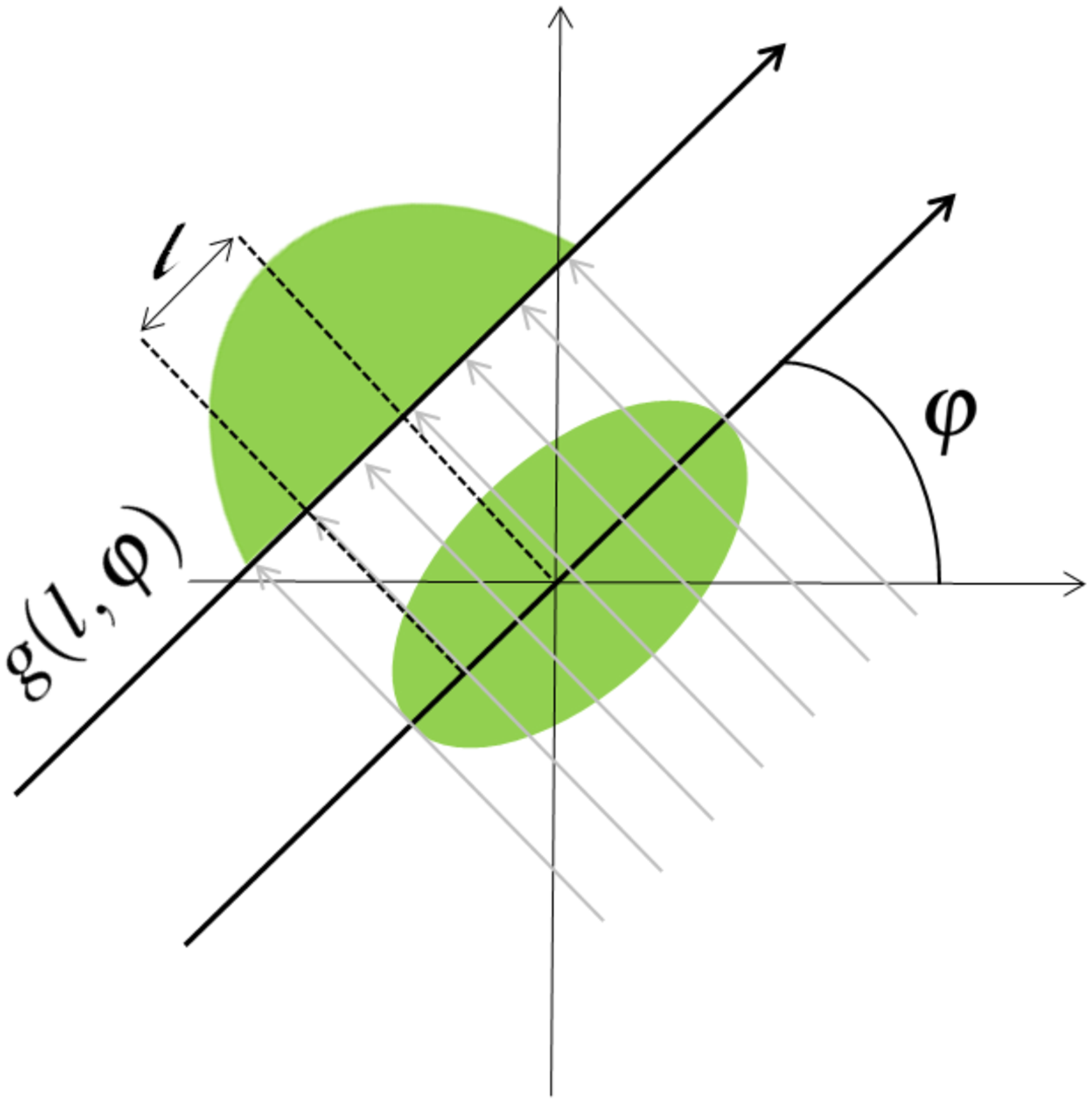}}
  \centerline{(A)}\medskip
\end{minipage}
\hfill
%\begin{minipage}[b]{0.3\linewidth}
%  \centering
%  \centerline{\includegraphics[width=3.0cm]{FanBeam1.eps}}
%  \centerline{(B)}\medskip
%\end{minipage}
%\hfill
\begin{minipage}[b]{0.45\linewidth}
  \centering
  \centerline{\includegraphics[width=4.5cm]{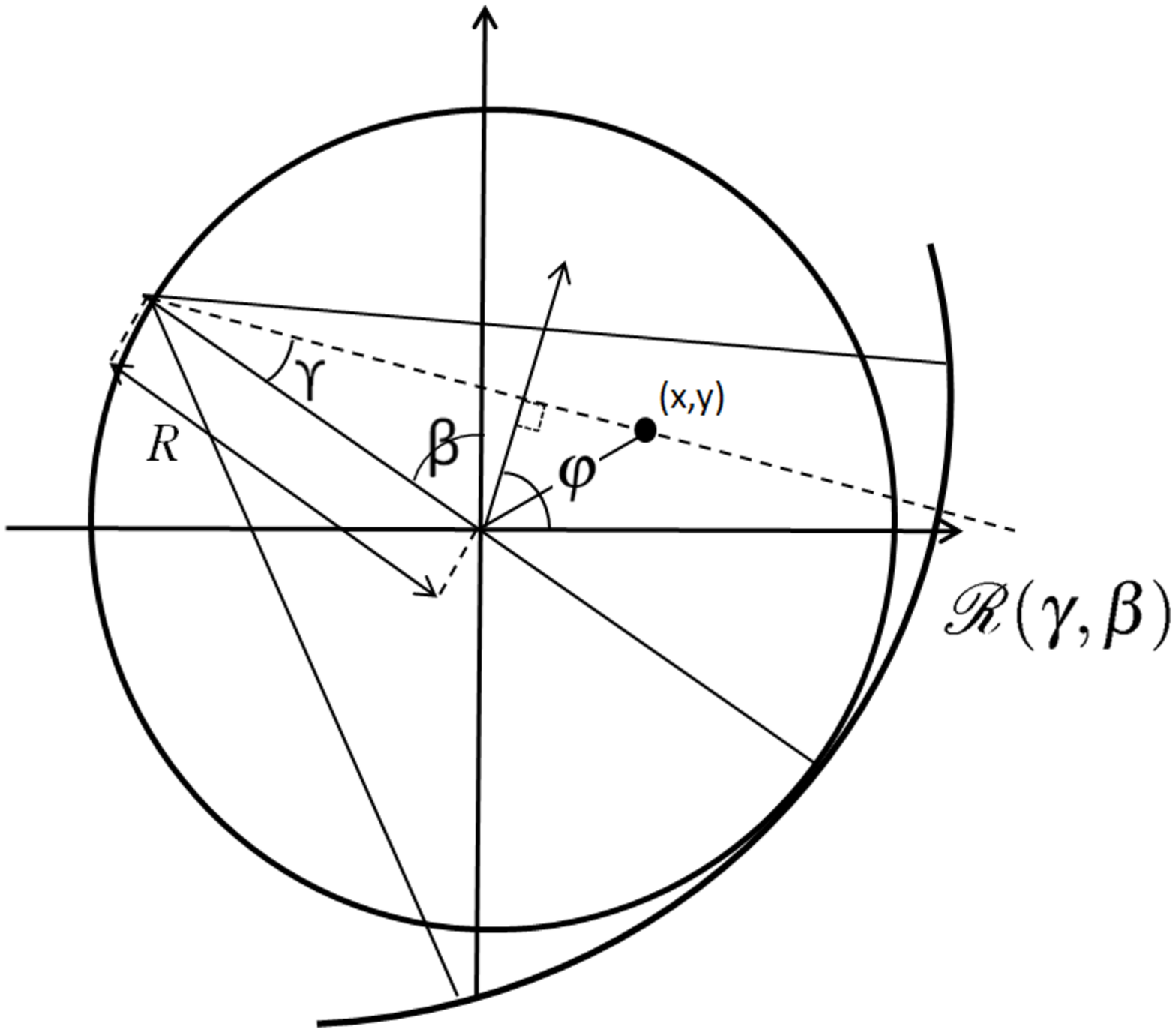}}
  \centerline{(B)}\medskip
\end{minipage}
\caption{(A) Parallel beam geometry and (B) $3^{rd}$ generation Fan beam geometry.} % with a flat detector (C) Fan beam geometry, $3^{rd}$ generation.}
\label{fig:fanpara}
\end{figure}

\begin{figure}[htb]
\begin{minipage}[b]{.48\linewidth}
  \centering
  \centerline{\includegraphics[width=4.0cm]{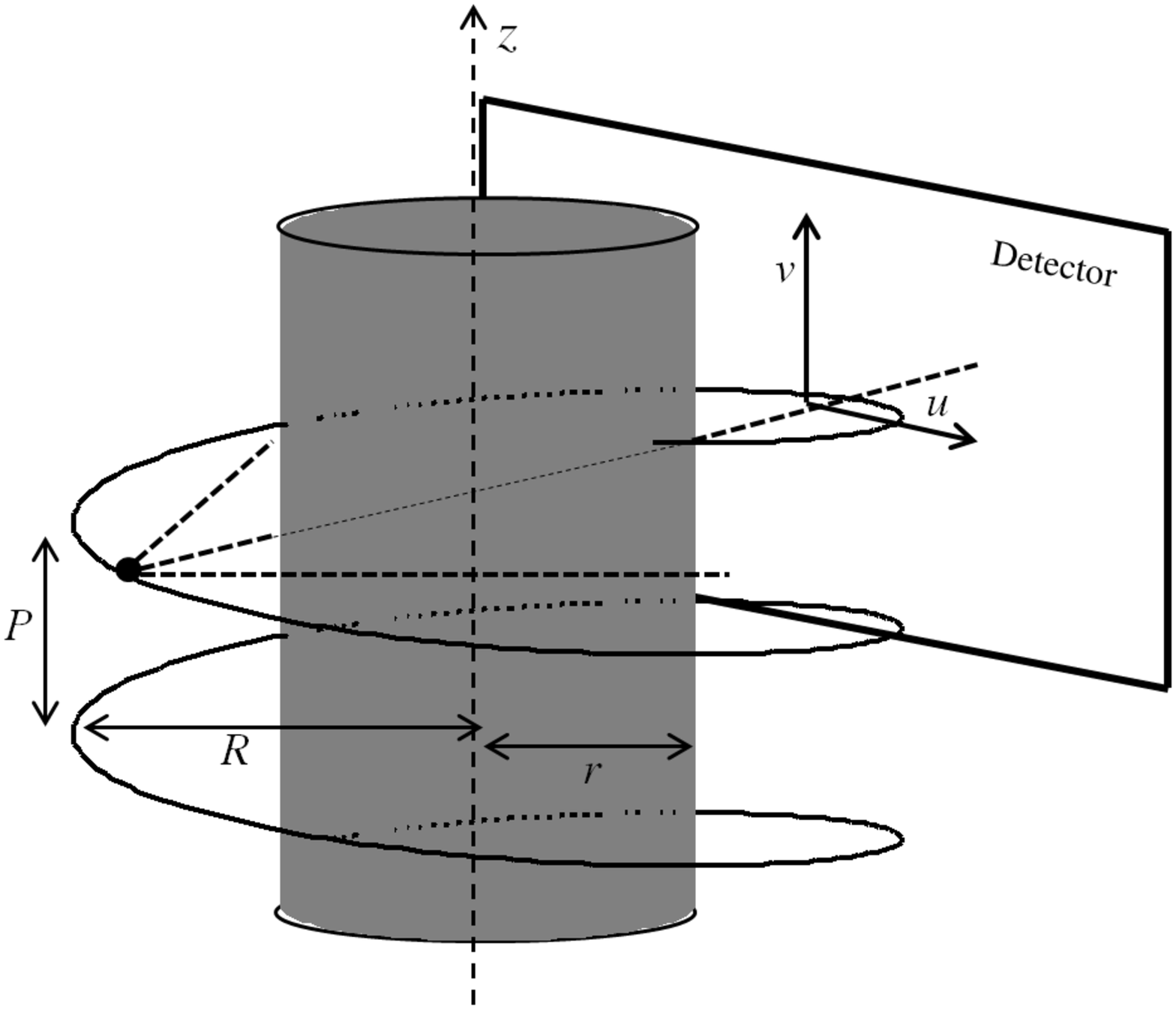}}
  \centerline{(A)}\medskip
\end{minipage}
\hfill
\begin{minipage}[b]{0.48\linewidth}
  \centering
  \centerline{\includegraphics[width=4.0cm]{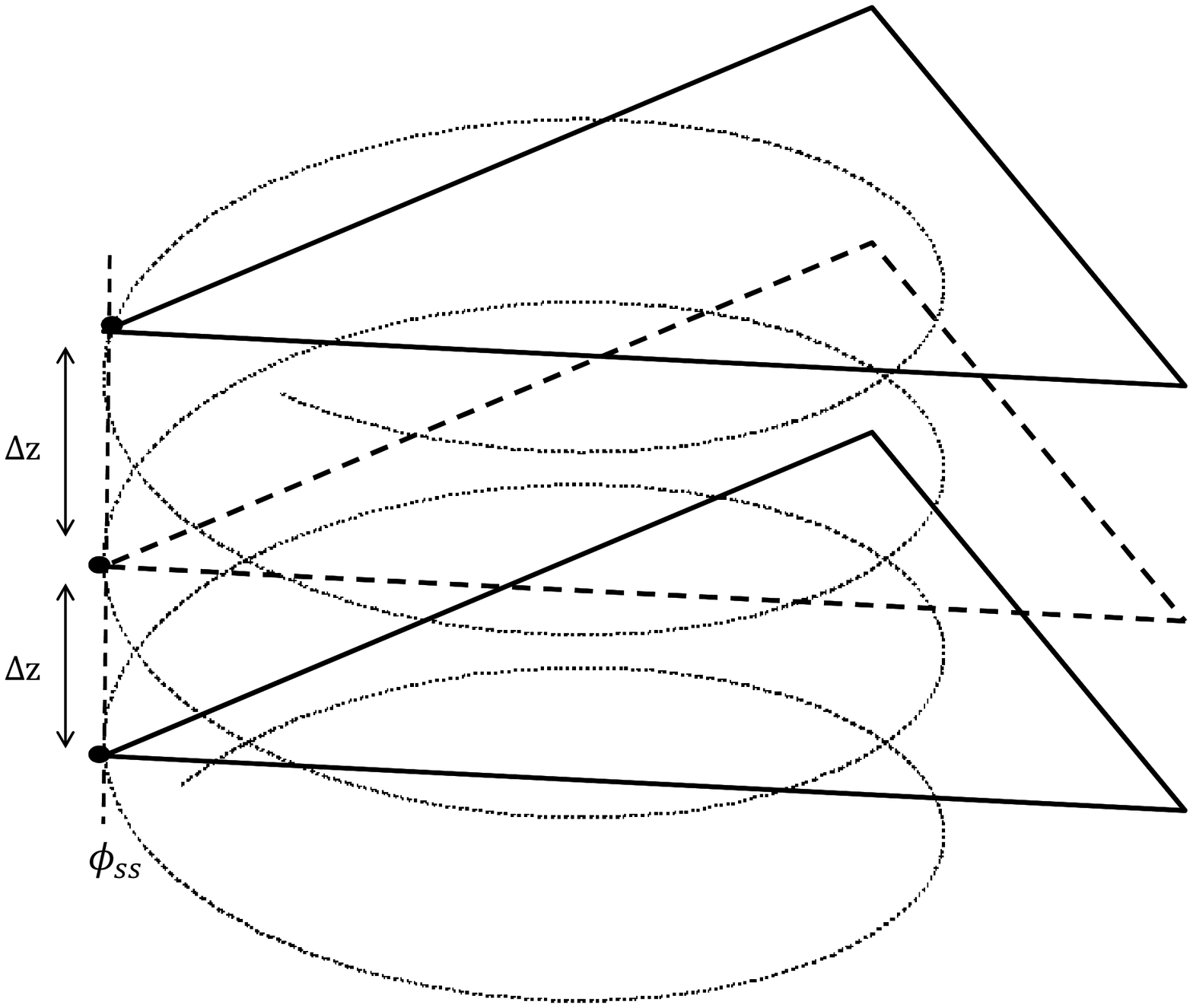}}
  \centerline{(B)}\medskip
\end{minipage}
\caption{(A) Helical trajectory and (B) the fan beams in parallel z-slices.}
\label{fig:helicalpara}
\end{figure}

To be able to use the central slice theorem and direct Fourier reconstruction in fan beam and helical geometries, the projections should be rearranged to parallel rays.
This redistribution of the rays is called rebinning \cite{Besson99,SSRB}. Since we use PPFT as our measurement matrix \mbf{A}, all the parallel rays
should be placed on equally sloped radial lines, $\varphi$ in (\ref{eqn:ppftangles}).
\\ \\
{\small\textbf{\emph{1- Fan beam to Equally Sloped Parallel Beams:}}}\\
Two interpolation steps are needed for fan beam geometry. In the first step, projections are interpolated
on  equally sloped radial lines, as shown in Figure \ref{fig:BHBV}-A. This step makes use of the following relationships between
fan and parallel beams:
\begin{eqnarray}\label{eqn:fanrebin}
\mathcal{R}(\gamma,\beta) &=& g(R\sin\gamma,\beta+\gamma)  \nonumber \\
l &=& R\sin\gamma \nonumber \\
\varphi &=& \beta+\gamma
\end{eqnarray}
where $\gamma$, $R$, $\varphi$ and $\beta$ are geometry parameters defined in Figure \ref{fig:fanpara}.
$\mathcal{R}(\gamma,\beta)$ is the fan beam projected data and $g(R\sin\gamma,\beta+\gamma)$ is the
corresponding rebinned parallel ray. These radial lines are then zero-padded and
the 1D Fourier transforms of the zero-padded radial lines are calculated. This is equivalent to oversampling in the Fourier domain.
In the second interpolation step, the oversampled radial Fourier coefficients are interpolated on pseudo polar grids, shown in Figure \ref{fig:BHBV}-B.
Since  the radial coefficients are oversampled, the interpolation error in this step is manageably small. The output of this step is the measured data \mbf{y}.
% ======================================================================================================================
%                                                       rebinning helical
% ======================================================================================================================
\\ \\
{\small \textbf{\emph{2- Helical Geometry to Equally Sloped Beams:}}}\\
To reconstruct the helically scanned objects, the scanned cone beam data are first converted to fan-beam data and then the fan beam data are converted to parallel beams.
This rebinning process is based on the method introduced in \cite{SSRB}, called
Cone Beam Single Slice ReBinning (CB-SSRB). CB-SSRB consists of the following two steps:
\begin{enumerate}
 \item For each source position in the helical trajectory, $\psi$, fix the z-sampling distances.
 \item For each z-slice, calculate the complete fan-beam set, from which the image can be estimated. This step uses the following
 equation to interpolate the cone beam scanned data on the fan beam points of interest:
\begin{eqnarray}
p_z(\varphi,u) &\simeq& \frac{\sqrt{u^2+D^2}}{\sqrt{u^2+v^2+D^2}}g_{\psi}(u,v), \nonumber \\
 v &=& \frac{u^2+D^2}{RD} \Delta z
\end{eqnarray}
where $p_z(\varphi,u)$ is estimated fan beam projection at source angle $\varphi$ and axial position $z$,
$g_{\psi}(u,v)$ is the cone beam projections at helical position, $D$ is the distance between the source
and the origin of the detector, and $u$, $v$, and $R$ are geometry parameters defined in Figure \ref{fig:helicalpara}.
Each interpolated fan beam projection is weighted by:
\end{enumerate}
\[ w(\phi_{ss},u)=
\left
\{
    \begin{array}{ll}
      \sin^2(\frac{\pi \phi_{ss}}{2(2\gamma_T+2\gamma)}) & \text{if} \phi_{ss} \in[0,2\gamma_T+2\gamma] \\
      1 & \text{if} \phi_{ss} \in[2\gamma_T+2\gamma, \pi+2\gamma] \\
      \sin^2(\frac{\pi (\pi+2\gamma_T-\phi_{ss})}{2(2\gamma_T-2\gamma)}) & \text{if} \phi_{ss} \in[\pi+2\gamma,\pi+2\gamma_T]
    \end{array}
\right. \]
where $\phi_{ss}=(\dfrac{\pi}{2}+\gamma_T)(1-\dfrac{\Delta z}{d})$, $d = 0.5P(\pi/2+\gamma_T)/(2\pi)$, $P$ is the pitch of the helical trajectory, and $2\gamma_T$ is the maximum fan angle.
The parallel beams $g(l,\varphi)$ are estimated from the weighted fan beams $p_z(\varphi, u)$ using (\ref{eqn:fanrebin}), from which \mbf{y} will be calculated
by computing the 1D Fourier transform of $g(l,\varphi)$s.
% ======================================================================================================================
%                                                       Confidence Matrix
% ======================================================================================================================
\subsection{Generalizing the CS Model to Adapt to Nonuniform Measurement Noise and Interpolation Error}\label{sec:confmatrix}
\begin{figure}[htb]
\begin{minipage}[b]{.98\linewidth}
\centering
  \centerline{\includegraphics[width=9.0cm]{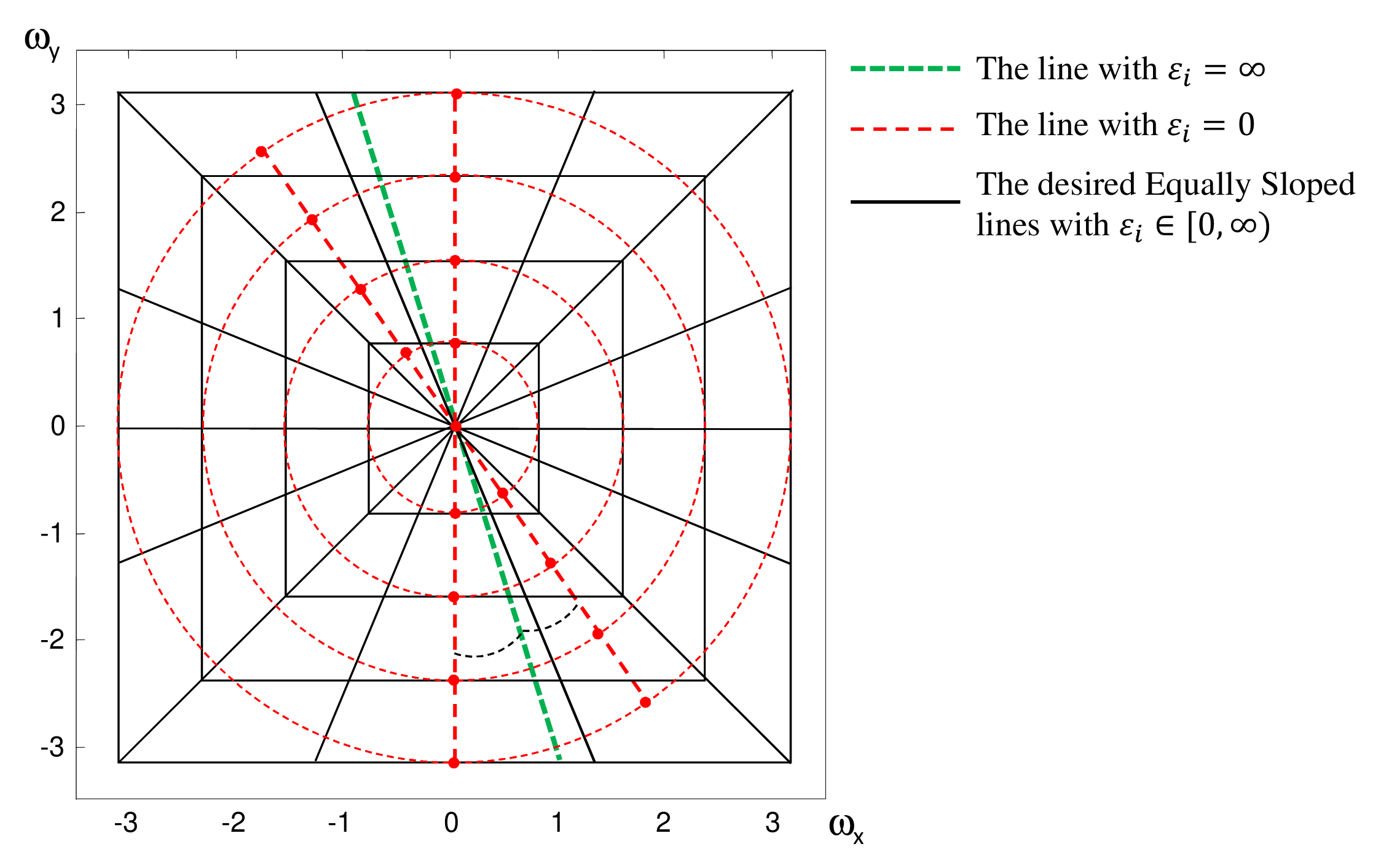}}
\end{minipage}
\caption{Calculating the effect of the interpolation error for inclusion in the Error Adaptation Weight (EAW).}
\label{fig:confidence}
\end{figure}

In section \ref{sec:MAP} we showed the MAP estimator of CT is a form of weighted CS  (\ref{eqn:modelL1} and \ref{eqn:modelTV}),
in which the weight is a function of noise variance and is denoted as $D$ in (\ref{eqn:di}).
We assert that the effects of noise variance and
interpolation error can be lumped together into the form of an \emph{Error Adaptation Weight (AEW)},
denoted by \mbf{c}:
\begin{equation} \label{eqn:ci}
d_i = \frac{1}{\sigma_{y_i}} \longrightarrow c_i = \frac{1}{\sigma_{y_i}+e_i}
\end{equation}
in which $e_i$ is the interpolation error. The greater the interpolation error, the greater the uncertainty about the
value of $\bar{\mbf{y}}_i$, so the effect of interpolation error is similar to the effect of the measurement noise $\sigma_{y_i}$.
AEW can be rewritten as:
\begin{equation} \label{eqn:ci2}
c_i = \frac{1}{\sigma_{y_i}+\epsilon_i \sigma_{y_i}} = \frac{1}{\sigma_{y_i}} \times \frac{1}{1+\epsilon_i}
\end{equation}
Using this definition, the generalized CS is as follows:
\begin{eqnarray}
\label{eqn:BPC}
 \hat{\mbf{x}} &=& \underset{\mrm{x}}{\operatorname{argmin}}\,\frac{1}{2}\|\mbf{c} \bullet (\mbf{y}- \mbf{Ax})\|^2_2+\lambda_{R_1} \|W^T \mbf{x}\|_q + \lambda_{R_2} \mathbf{TV}(f), \nonumber \\
 \mbf{c} &=& \left[ \begin{array}{c} c_1 \\ \vdots \\ c_{(k  n_\varphi) \times n^2} \end{array} \right] \propto vec(D) \bullet \frac{1}{1+\mbf{\epsilon}}
\end{eqnarray}
where $\bullet$ denotes the element-by-element multiplication, $vec(.)$ converts the matrix into a column vector,
and $\mbf{\epsilon}=[\epsilon_1,...,\epsilon_{kn_{\varphi}\times n^2}], \epsilon_i \in [0,\infty)$ represents the effect of interpolation error.\\
The method used for calculating $\epsilon$ is illustrated in Figure \ref{fig:confidence}.
The value of $\epsilon_i$ for a line exactly between two polar lines is $\infty$, since its distance from the true measured values are maximal and therefore the error is maximal.
Using (\ref{eqn:ci}) this can be thought as $e_i\rightarrow \infty$ and consequently $\epsilon_i \rightarrow \infty$ or $c_i \rightarrow 0$.
The closer the equally sloped lines are to the rays on which the measurements are done, the interpolation error gets smaller
and $\epsilon_i$'s on that line get closer to zero. Finally, if the desired equally sloped rays are exactly on the polar lines,
the interpolation error $e_i$ is zero which is equivalent to $\epsilon_i = 0$. \\

%--------------------------------------------------------------
%			Efficient Reformulation of CT model
%--------------------------------------------------------------
\section{EM solution for the generalized CS model} \label{sec:EM-MAP}
It has previously been shown that the quality of the reconstructed image can be improved by combining the sparsity and total variation penalty terms \cite{RecPF}
both of which are incorporated in (\ref{eqn:BPC}). It is thus the equation used
in this paper to recover  CT images from  undersampled data. In
 solving this optimization problem, we use the
Fast Composite Splitting Algorithm (FCSA) \cite{FCSA} to decompose (\ref{eqn:BPC}) into two simpler sub-problems similar to (\ref{eqn:modelTV}) and (\ref{eqn:modelL1}), which are
given by:
\begin{eqnarray}
\label{eqn:subproblems}
 \hat{\mbf{x}}_1 &=& \underset{\mrm{x}}{\operatorname{argmin}}\,\frac{1}{2}\|\mbf{c} \bullet (\mbf{y}- \mbf{Ax})\|^2_2+\lambda_{R_1} \|W^T \mbf{x}\|_q \nonumber \\
 \hat{\mbf{x}}_2 &=& \underset{\mrm{x}}{\operatorname{argmin}}\,\frac{1}{2}\|\mbf{c} \bullet (\mbf{y}- \mbf{Ax})\|^2_2+\lambda_{R_2} \mathbf{TV}(f)
\end{eqnarray}
Calculating $\hat{\mbf{x}}_1$ and $\hat{\mbf{x}}_2$, FCSA proposes that the solution to the problem can be obtained by a linear combination of the
solutions of the two sub-problems, \emph{i.e.},
\begin{equation}\label{eqn:FCSA_linear}
  \hat{\mbf{x}} = \delta \hat{\mbf{x}}_1 + (1-\delta) \hat{\mbf{x}}_2
\end{equation}
in which $\delta$ is a weight that is defined as a function of the value of the objective functions of the two subproblems, denoted as $f_1$ and $f_2$ and is
given by $\delta = f_2/(f_1+f_2)$.
Therefore, an FCSA based EM method is proposed to recover the CT images from the X-ray projections and is called the FCSA-LEM algorithm.
\subsection{Latent variable and EM method for $\ell_2-\ell_1$ and $\ell_2$-TV Optimization}
Here an efficient method is proposed to solve $\ell_2-\ell_1$ and $\ell_2$-TV subproblems in (\ref{eqn:subproblems}).
An Expectation-Maximization algorithm is used to solve these two optimizations problems \cite{ECME,EMWavelet} for a CT modeled by (\ref{eqn:GausModel}).
A latent variable $\mbf{z}$ is defined such that the problem in (\ref{eqn:GausModel}) can be written as\footnote{The definition and the effect of this latent variable in the final solution is similar to the variable splitting strategy used in alternating direction methods (ADM) \cite{ADM1,RecPF}.}:
\begin{equation}\label{eqn:hidden1}
  \mbf{y} = \mbf{A} \mbf{z} + n_1
\end{equation}
in which $\mbf{z}$ is chosen to be:
\begin{equation}\label{eqn:hidden2}
  \mbf{z} = \mbf{x} + \alpha n_2 = W \theta + \alpha n_2
\end{equation}
The noise is split into two parts: $\mbf{n}=\alpha \mbf{A} n_1 + n_2$ and $p(n_1) = N(n_1;0,I)$,
$p(n_2) = N(n_2;0,D^{-1}-\alpha^2\mbf{AA}^T)$. Using this notation, the EM algorithm is as follows:\\
\textbf{E-step:}
Compute the conditional expectation of the log likelihood, given the observed data and the current
estimate $\theta^{(t)}=W^T\mbf{x}^{(t)}$:
\begin{equation} \label{eqn:Estep}
 Q(\theta,\theta^{(t)}) = E[\log p(\mbf{y},\mbf{z}|\theta)|\mbf{y},\theta^{(t)}]
\end{equation}
\textbf{M-Step:}
Update the estimate:
\begin{equation}\label{eqn:Mstep}
 Q^{(t+1)} = arg \max_{\theta} \{Q(\theta,\theta^{(t)})-h(\theta)\}
\end{equation}
In the E-step, $\mbf{z}^{(t)} = E[\mbf{z}|\mbf{y},\theta^{(t)}]$ should be calculated and plugged in (\ref{eqn:Estep}), in which
we need to calculate the likelihood $p(\mbf{y},\mbf{z}|\theta)=p(\mbf{y}|\mbf{z},\theta)p(\mbf{z}|\theta)=p(\mbf{y}|\mbf{z})p(\mbf{z}|\theta)$.
Since $p(\mbf{y}|\mbf{z}) \sim N(\mbf{y};\mbf{Az},D^{-1}-\alpha^2\mbf{AA}^T)$ and $p(\mbf{z}|\theta) \sim N(\mbf{z};0,\alpha^2I)$,
$\log p(\mbf{y},\mbf{z}|\theta)=-\frac{\|\mbf{x}-\mbf{z}\|^2}{2\alpha^2}+K = -\frac{\mbf{x}^T\mbf{x}-2\mbf{x}^T\mbf{z}}{2\alpha^2}+K'$ and
therefore we have the following equation for $Q(\theta,\theta^{(t)})$:
\begin{equation}\label{eqn:qtMstep}
  Q(\theta,\theta^{(t)})=-\frac{\|\mbf{x}-\mbf{z}^{(t)}\|^2}{2\alpha^2}+K = -\frac{\mbf{x}^T\mbf{x}-2\mbf{x}^T\mbf{z}^{(t)}}{2\alpha^2}+K'
\end{equation}
in which $K$ and $K'$ do not depend on $\mbf{x}$ and as a result $\theta$.
Since $p(\mbf{z}|\mbf{y},\hat{\theta}^{(t)}) \propto p(\mbf{y}|\mbf{z})p(\mbf{z}|\hat{\theta}^{(t)})$, in which both
$p(\mbf{y}|\mbf{z})$ and $p(\mbf{z}|\hat{\theta}^{(t)})$ are Gaussian,
$p(\mbf{z}|\mbf{y},\hat{\theta}^{(t)})$ is Gaussian with mean value of $\mbf{z}^{(t)}=E[p(\mbf{z}|\mbf{y},\hat{\theta}^{(t)})]$ \cite{Statistic1,Statistic2}:
\begin{eqnarray}
 \mbf{z}^{(t)} &=& \mbf{x}^{(t)} + C_{\mbf{z}} \mbf{A}^T (\mbf{A}C_\mbf{z}\mbf{A}^T+C_{n_2})^{-1}(y-\mbf{A}\mbf{x}^{(t)})
\end{eqnarray}
in which $C_{n_2}=D^{-1}-\alpha^2\mbf{AA}^T$ and $C_{\mbf{z}} = \alpha^2I$.
Therefore, the E-step can be summarized by the calculation of:
\begin{eqnarray}\label{eqn:Estep3}
 \mbf{z}^{(t)} &=& \mbf{x}^{(t)} + \alpha^2 \mbf{A}^T D (y-\mbf{A}x^{(t)}) \nonumber \\
 &=& W \theta^{(t)} + \alpha^2 \mbf{A}^T D (y-\mbf{A}W \theta^{(t)})
\end{eqnarray}
By inclusion of the rebinning error, \emph{i.e.} using the EAW introduced in (\ref{eqn:BPC}), this step will be:
\begin{eqnarray}\label{eqn:Estep2}
 \mbf{z}^{(t)} &=& \mbf{x}^{(t)} + \alpha^2 \mbf{A}^T (\mbf{c} \bullet y- \mbf{c} \bullet \mbf{A}x^{(t)}) \nonumber \\
 &=& W \theta^{(t)} + \alpha^2 \mbf{A}^T (\mbf{c} \bullet y- \mbf{c} \bullet \mbf{A}W \theta^{(t)})
\end{eqnarray}
that will be followed by an M-step:
\begin{eqnarray}\label{eqn:Mstep2}
\theta^{(t+1)} &=& \underset{\mrm{\theta}}{\operatorname{argmax}}\, \{ -\frac{\|W\theta-\mbf{z}^{(t)}\|^2}{2\alpha^2}-h(\theta) \} \nonumber \\
               &=& \underset{\mrm{\theta}}{\operatorname{argmin}}\, \{  \frac{\|\theta-W^T\mbf{z}^{(t)}\|^2}{2\alpha^2}+h(\theta) \} \nonumber \\
\mbf{x}^{(t+1)} &=& \underset{\mrm{x}}{\operatorname{argmin}}\, \{ \frac{\|x-\mbf{z}^{(t)}\|^2}{2\alpha^2}+h(x) \}
\end{eqnarray}
Equation (\ref{eqn:Mstep2}) has a closed form solution for some special cases, e.g. soft thresholding if $h(\theta) = \lambda_R \|\theta\|_1$ \cite{WaveThresh} and the TV denoising problem if $h(\theta) = \lambda_R \mbf{TV}(W\theta)$ \cite{TVmin}.

\subsection{Solving the Generalized CS Model}
The pseudocode shown in Algorithm \ref{alg:FCSA} outlines the FCSA-LEM method used to solve the
generalized CS problem derived in section \ref{sec:confmatrix}, using the method introduced in (\ref{eqn:Estep3})-(\ref{eqn:Mstep2}).
In this algorithm $prox_L\{g(x),z\} = \underset{\mrm{x}}{\operatorname{argmin}}\, g(x)+\frac{L}{2}\|x-z\|_2^2$.
The optimization problem in step 2 of Algorithm \ref{alg:FCSA}:
\begin{equation*}
 \hat{\mbf{x}}_1  = W(prox_{1/\alpha^2} \{\lambda_{R_1} \|\theta\|_1,W^T \mbf{z}\}
\end{equation*}
has a closed form solution given by:
\begin{equation}
  \hat{\mbf{x}}_1  =  W (\operatorname{sign}(W^T \mbf{z}) \max\{0,|W^T \mbf{z}|-\lambda_{R_1} \alpha^2\})
\end{equation}
To calculate $\hat{\mbf{x}}_2$ in step 3, a total variation minimization scheme is used and solved by a split Bregman based method \cite{TVmin}
\footnote{The final steps are very similar to iterative soft thresholding based methods \cite{TwIST}.}.
%-----------------------------------------------------------------------------
\begin{algorithm}[h!]
\caption{\small Pseudocode of the {\bf FCSA-LEM} algorithm used to solve the optimization problem.}
\begin{code}
{\bf Initialize:} $\alpha$, $\lambda_{R_1}$, $\lambda_{R_2}$, $\mbf{c}$, $r_1=0$, $t_1=1$, maxiter, tol \\
\\
 \>\uwhile $\frac{\|\hat{x}_k-\hat{x}_{k-1}\|_2}{\|\hat{x}_k\|}> \, $tol or $k <$ maxiter \udo\\
\uln \>\> $\mbf{z}=r_{k} + \alpha^2 \mbf{A}^T (\mbf{c} \bullet \mbf{y}- \mbf{c} \bullet \mbf{A}r_k)$\\
\uln \>\> $\hat{\mbf{x}}_1 = W(prox_{1/\alpha^2}\{\lambda_{R_1}\|W^T \mbf{x}\|_1,W^T \mbf{z}\})$\\
\uln \>\> $\hat{\mbf{x}}_2 = prox_{1/\alpha^2}\{\lambda_{R_2}\mbf{TV}(\mbf{x}),\mbf{z}\}$\\
\uln \>\> $\hat{\mbf{x}}_k = \delta \hat{\mbf{x}}_1 + (1-\delta) \hat{\mbf{x}}_2$\\
\uln \>\> $t_{k+1} = \frac{1+\sqrt{1+4t_k^2}}{2}$ \\
\uln \>\> $r_{k+1} = \hat{\mbf{x}}_k + (\frac{t_{k}-1}{t_{k+1}})(\hat{\mbf{x}}_k-\hat{\mbf{x}}_{k-1})$ \\
\uln \>\> $k \leftarrow k+1$ \\
 \>\uend \uwhile
\end{code}
\label{alg:FCSA}
\end{algorithm}
%-----------------------------------------------------------------------------

% ---------------------------------------------------------------------------
%                           results
% ---------------------------------------------------------------------------
\section{Results} \label{sec:results}
\begin{figure}[htb]
\begin{minipage}[b]{.48\linewidth}
  \centering
  \centerline{\includegraphics[width=4.0cm]{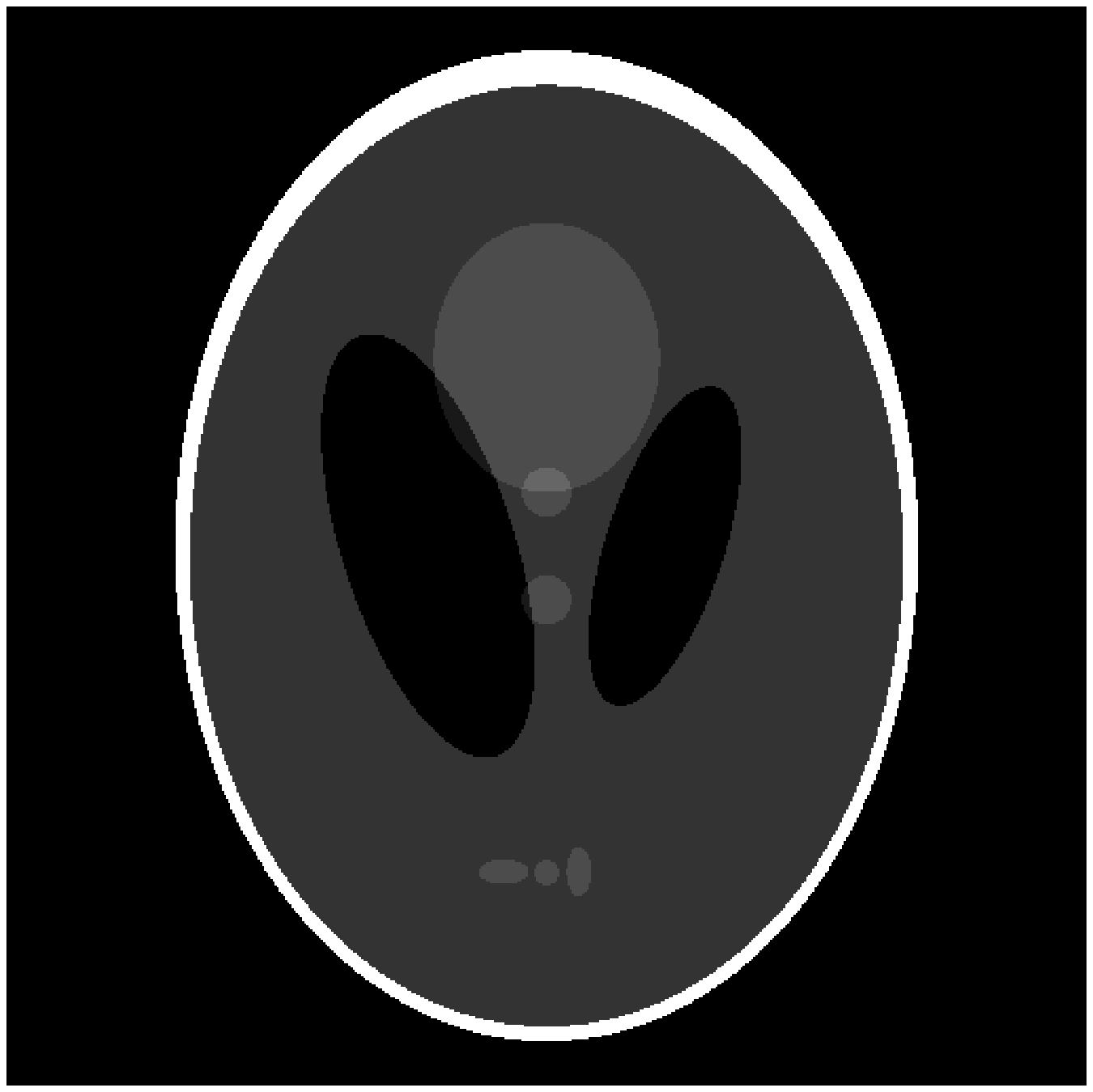}}
  \centerline{(A)}\medskip
\end{minipage}
\hfill
\begin{minipage}[b]{.48\linewidth}
  \centering
  \centerline{\includegraphics[width=4.0cm]{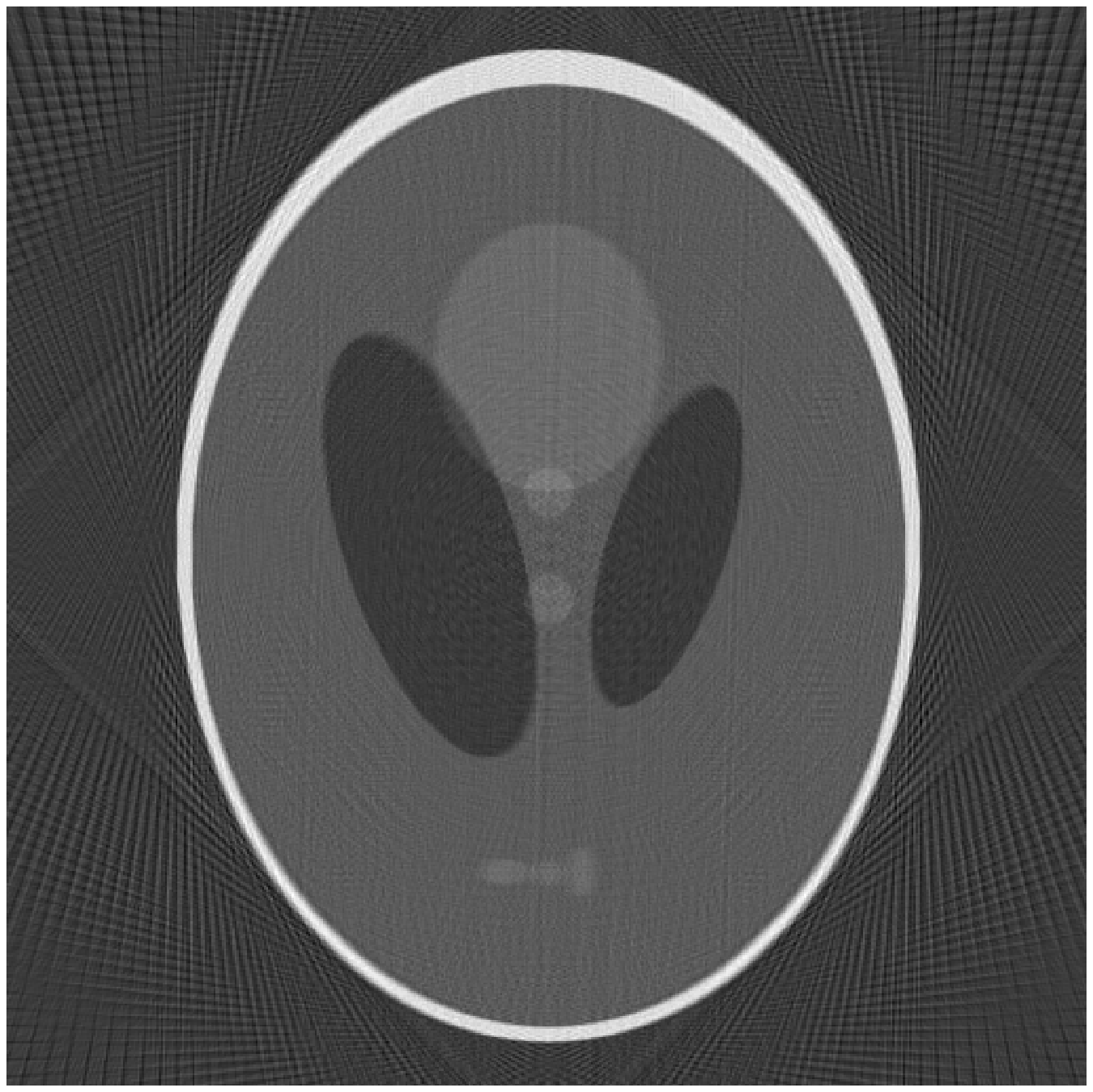}}
  \centerline{(B)}\medskip
\end{minipage}
\hfill
\begin{minipage}[b]{0.48\linewidth}
  \centering
  \centerline{\includegraphics[width=4.0cm]{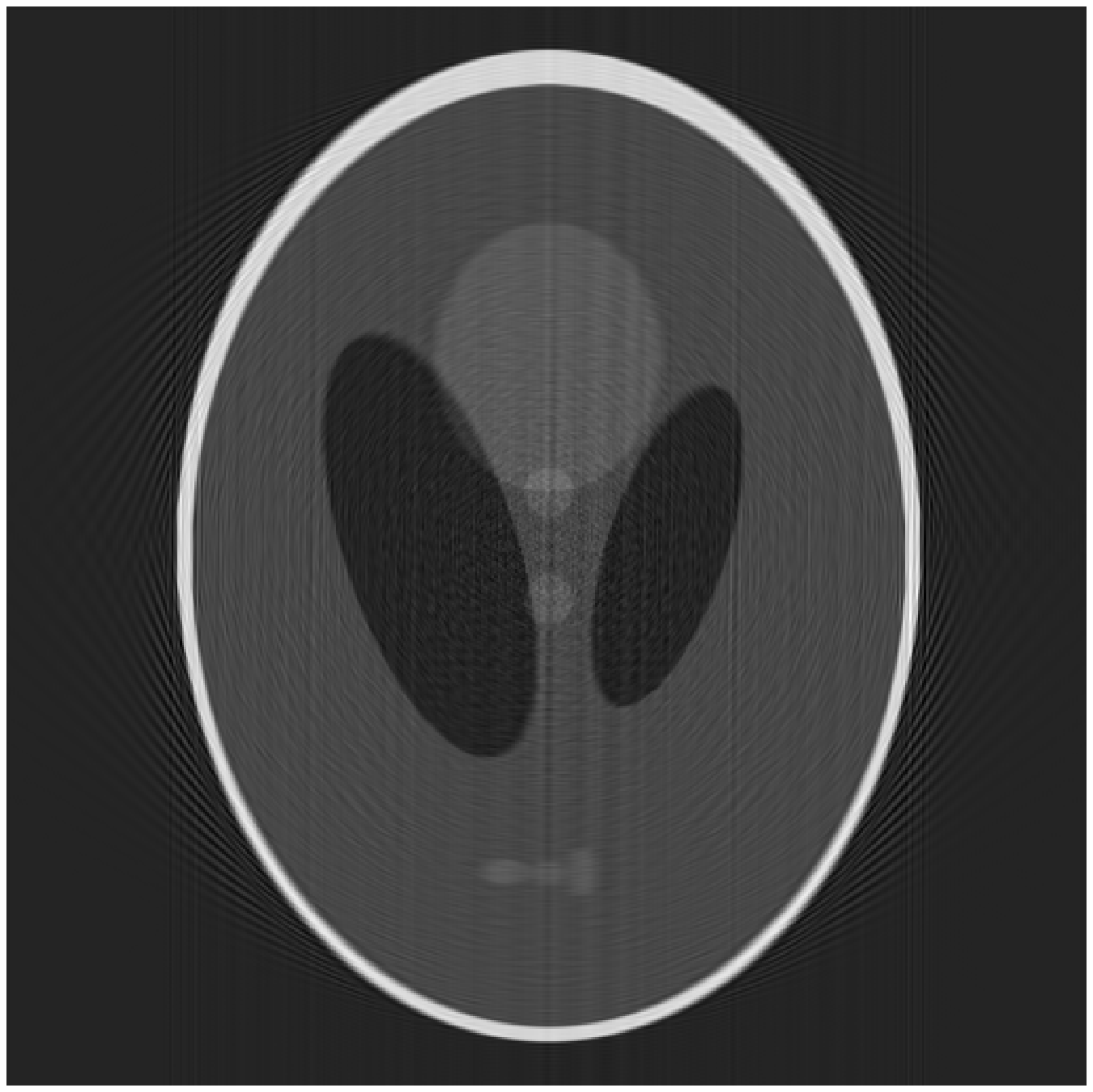}}
  \centerline{(C)}\medskip
\end{minipage}
\hfill
\begin{minipage}[b]{0.48\linewidth}
  \centering
  \centerline{\includegraphics[width=4.0cm]{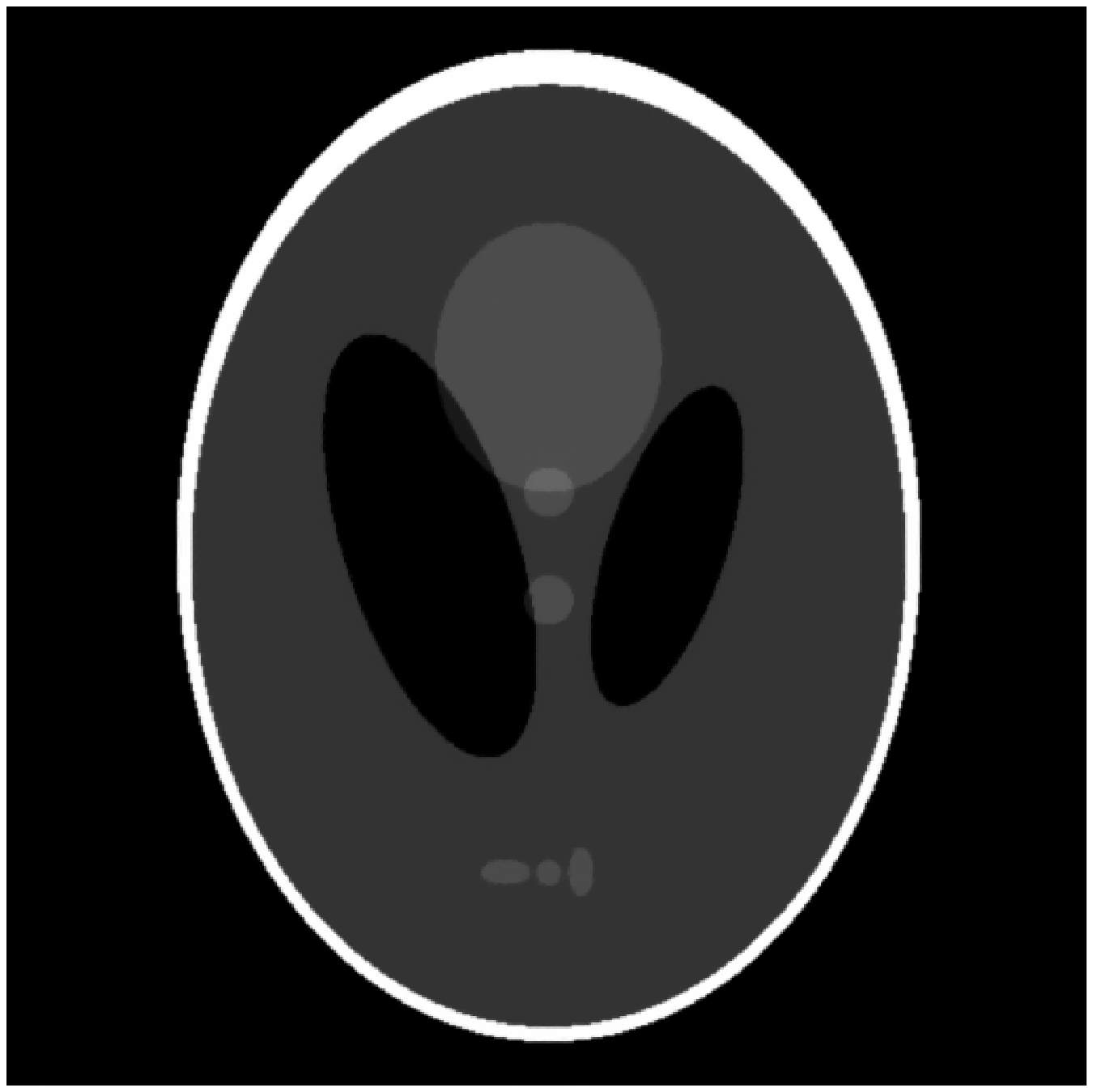}}
  \centerline{(D)}\medskip
\end{minipage}
\caption{(A) Original Shepp-Logan phantom Image.  Reconstructions using 128 projections with (B) inverse pseudo polar Fourier transform using the least squares
method (normalized error $\approx 0.9$), (C) an iterative soft thresholding based method (TwIST) (normalized error $\approx 10^{-1}$), and (D) the proposed
FCSA-LEM method (normalized error $\approx 10^{-2}$). The rebinned parallel rays are used in all three methods to recover the image.}
\label{fig:phan3methodcompare}
\end{figure}

\begin{figure}[htb!]
\begin{minipage}[b]{.9\linewidth}
  \centering
  \centerline{\includegraphics[width=9.0cm]{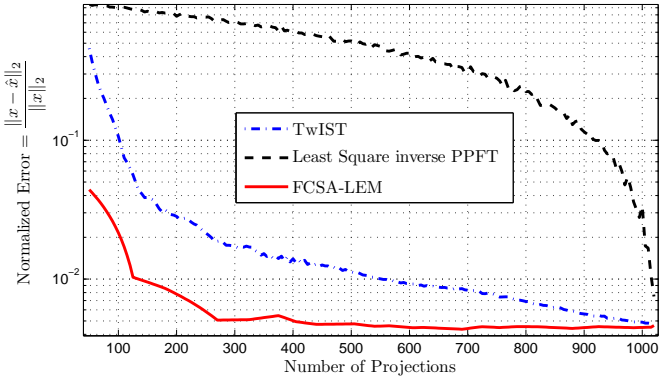}}
\end{minipage}
\caption{Normalized reconstruction error for the Shepp-Logan phantom  reconstructed with the inverse pseudo polar Fourier transform (PPFT) using
least squares (LS) method, using an iterative soft thresholding based method (TwIST), and the proposed FCSA-LEM method. Rebinned parallel rays
were used in all three methods to recover the image.}
\label{fig:NormalizedError_FBPvsCS}
\end{figure}

\begin{figure}[htb!]
\begin{minipage}[b]{.9\linewidth}
  \centering
  \centerline{\includegraphics[width=9.0cm]{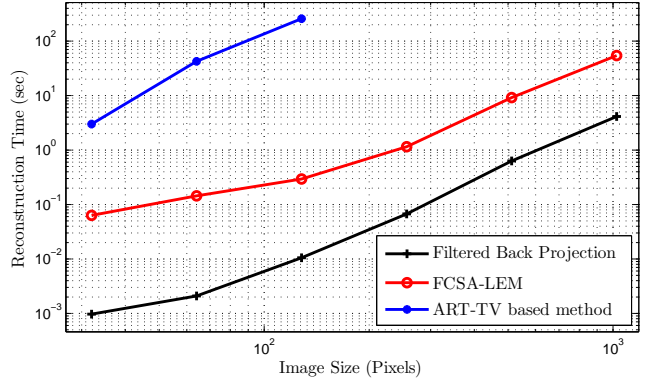}}
\end{minipage}
\caption{Reconstruction time comparison using a standard desktop computer, for: (1) fan beam filtered back projection (FBP) reconstruction,
(2) the proposed method (FCSA-LEM), and (3) a fan beam ART-TV based method.}
\label{fig:TimeFBPvsCSLog}
\end{figure}

\begin{figure}[htb!]
\begin{minipage}[b]{0.48\linewidth}
  \centering
  \centerline{\includegraphics[width=4.0cm]{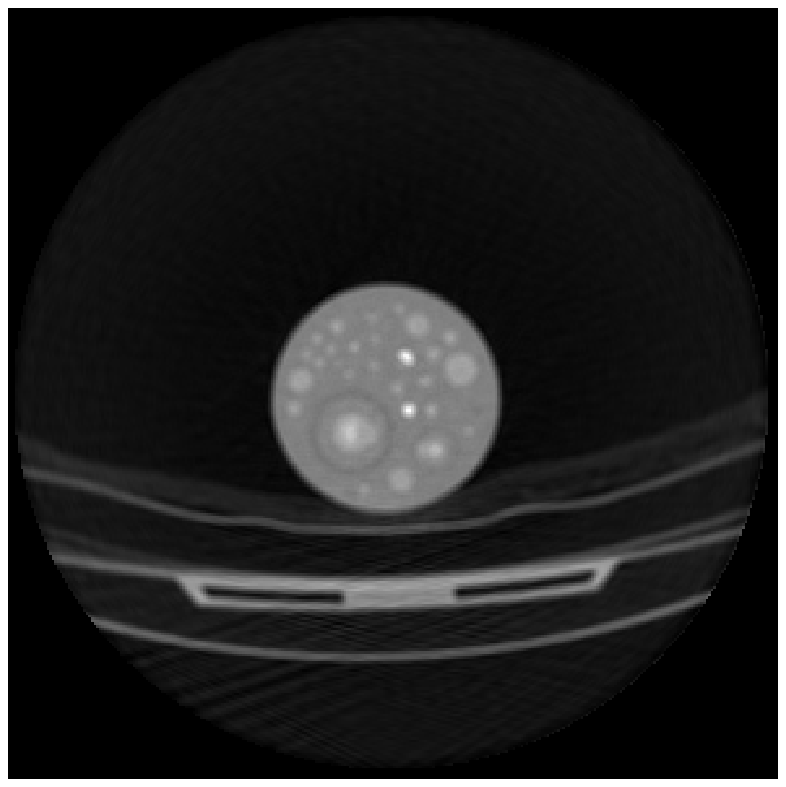}}
  \centerline{(A)}\medskip
\end{minipage}
\hfill
%\begin{minipage}[b]{0.3\linewidth}
%  \centering
%  \centerline{\includegraphics[width=3.0cm]{HeartPhantomLS}}
%  \centerline{(B)}\medskip
%\end{minipage}
%\hfill
\begin{minipage}[b]{0.48\linewidth}
  \centering
  \centerline{\includegraphics[width=4.0cm]{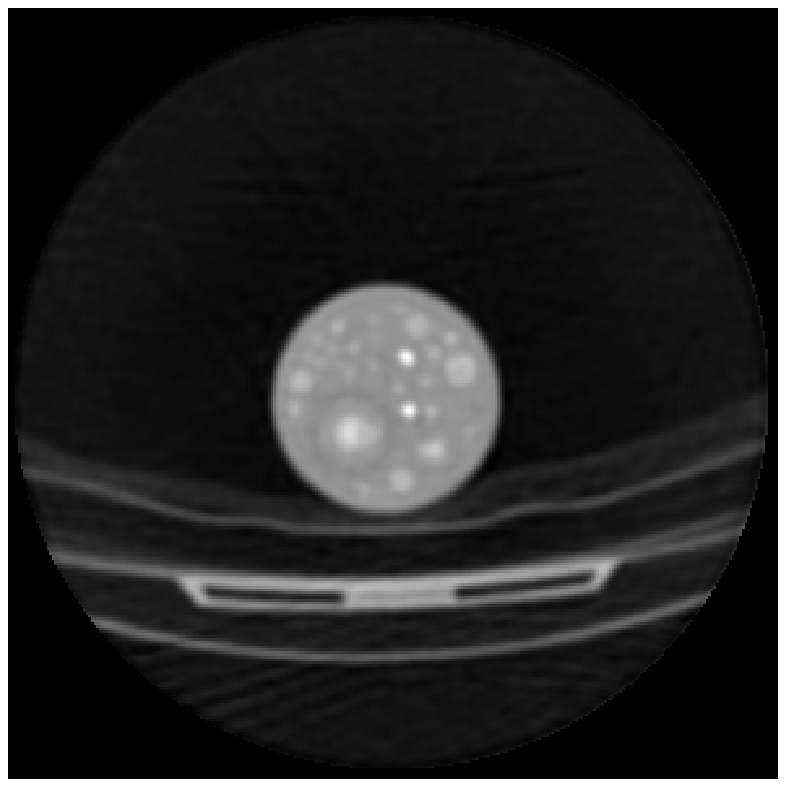}}
  \centerline{(B)}\medskip
\end{minipage}
\caption{Comparison of FBP and the proposed method for a cardiac plaque phantom made in Toronto General Hospital.
The scan protocol was 50mAs and 120kVp. (A) Image reconstructed from 1200 projections using FBP.
(B) Image reconstructed from 256 projections with the proposed method.}
\label{fig:phantest}
\end{figure}

\begin{figure}[htb!]
\begin{minipage}[b]{0.48\linewidth}
  \centering
  \centerline{\scalebox{-1}[1]{\includegraphics[width=4.0cm]{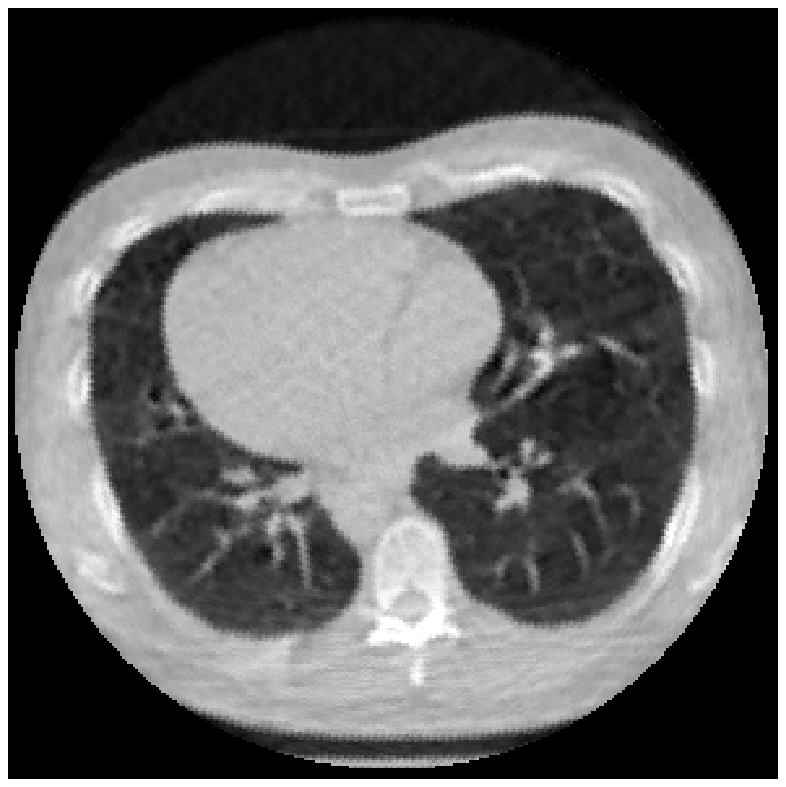}}}
  \centerline{(A)}\medskip
\end{minipage}
\hfill
\begin{minipage}[b]{0.48\linewidth}
  \centering
  \centerline{\scalebox{-1}[1]{\includegraphics[width=4.0cm]{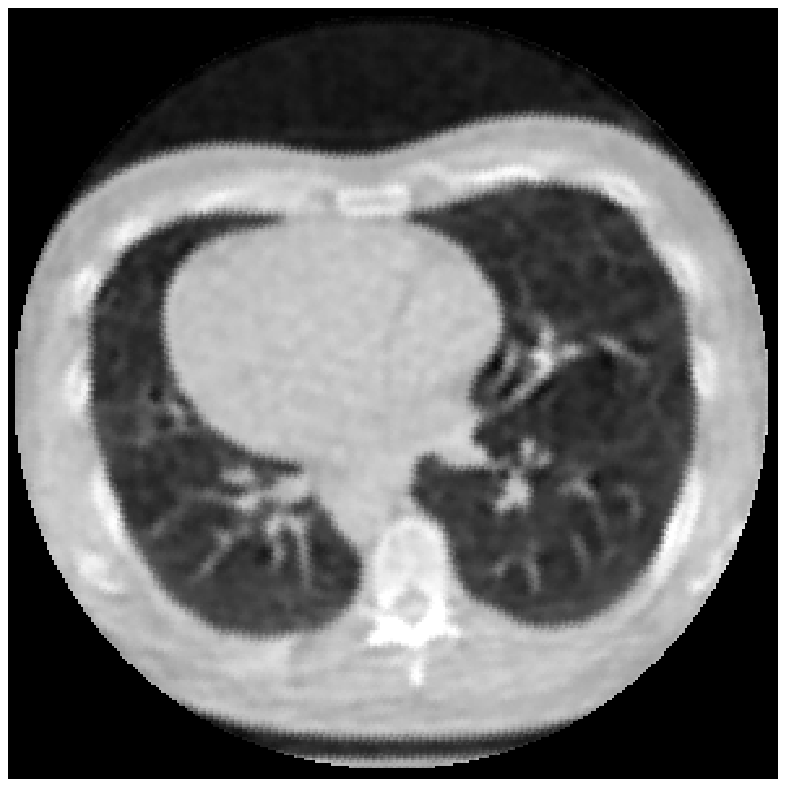}}}
  \centerline{(B)}\medskip
\end{minipage}
\caption{Comparison of FBP and the proposed method for the chest CT scan data  from a patient.
The scan protocol was 50mAs and 120kVp. (A) Image reconstructed from 1200 projections with FBP.
(B) Image reconstructed from 256 projections with the proposed method.}
\label{fig:patienttest}
\end{figure}

\begin{figure*}[htb!]
\begin{minipage}[b]{0.3\linewidth}
  \centering
  \centerline{\includegraphics[width=5.0cm]{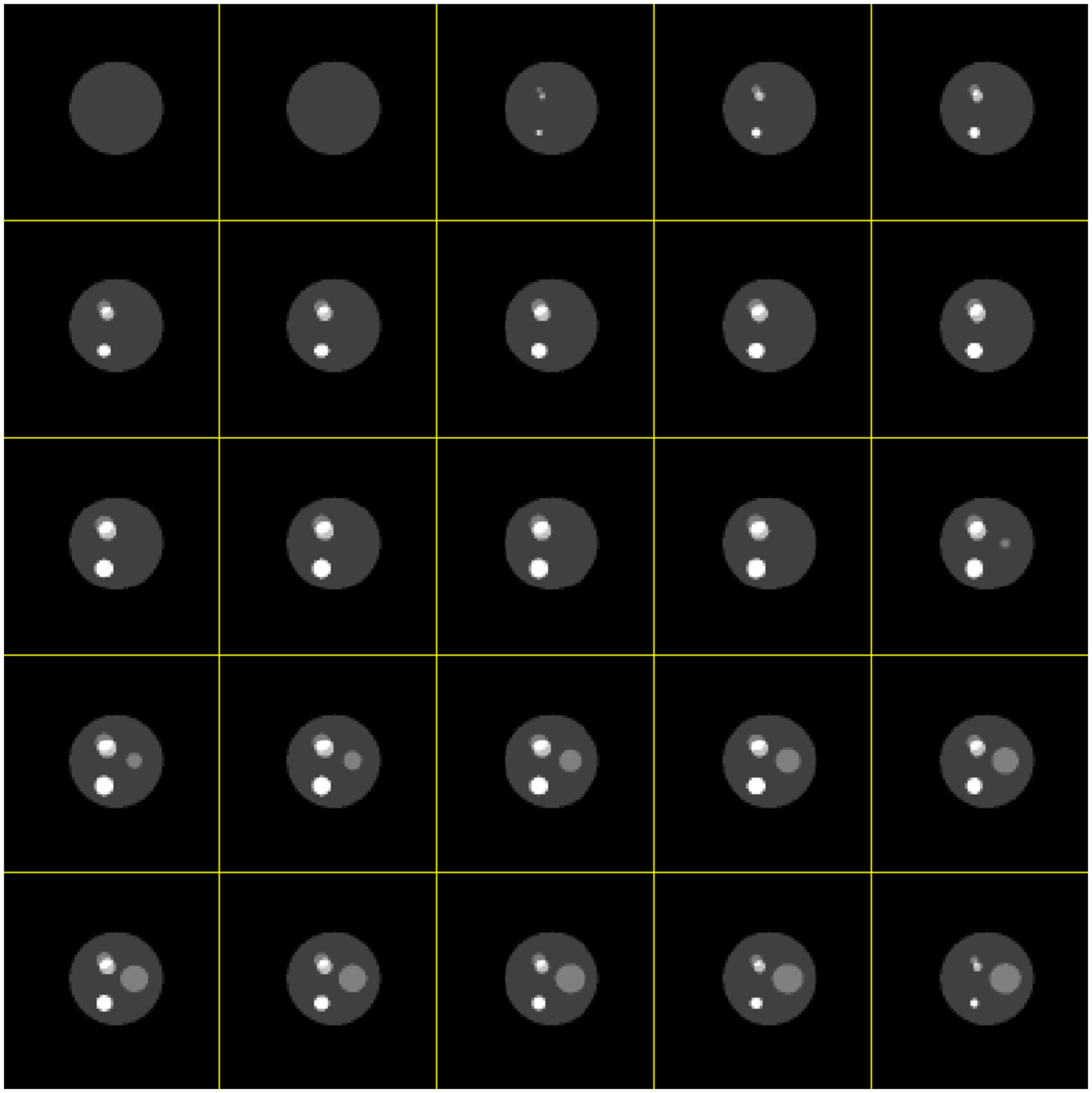}}
  \centerline{(A)}\medskip
\end{minipage}
\hfill
\begin{minipage}[b]{0.3\linewidth}
  \centering
  \centerline{\includegraphics[width=5.0cm]{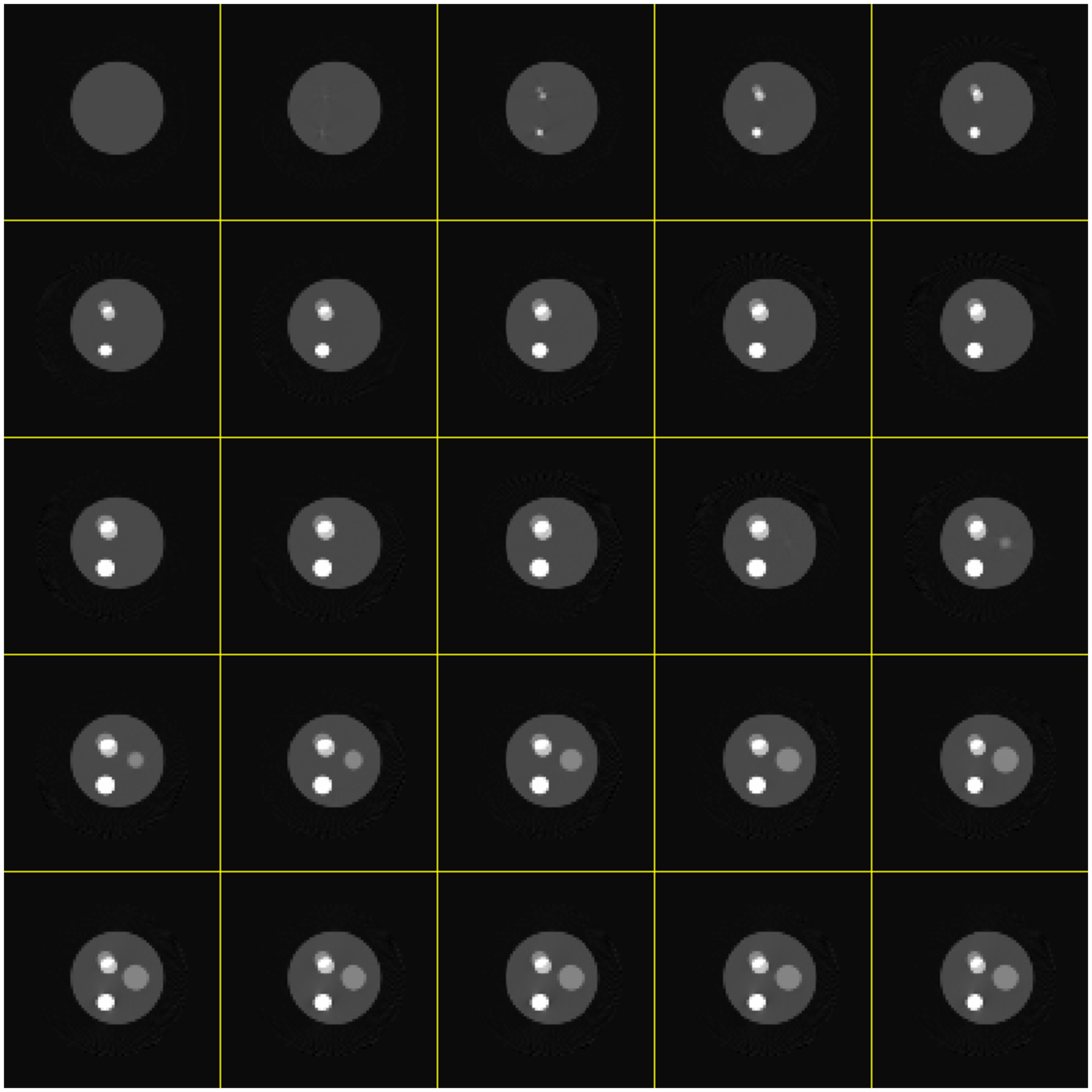}}
  \centerline{(B)}\medskip
\end{minipage}
\hfill
\begin{minipage}[b]{0.3\linewidth}
  \centering
  \centerline{\includegraphics[width=5.0cm]{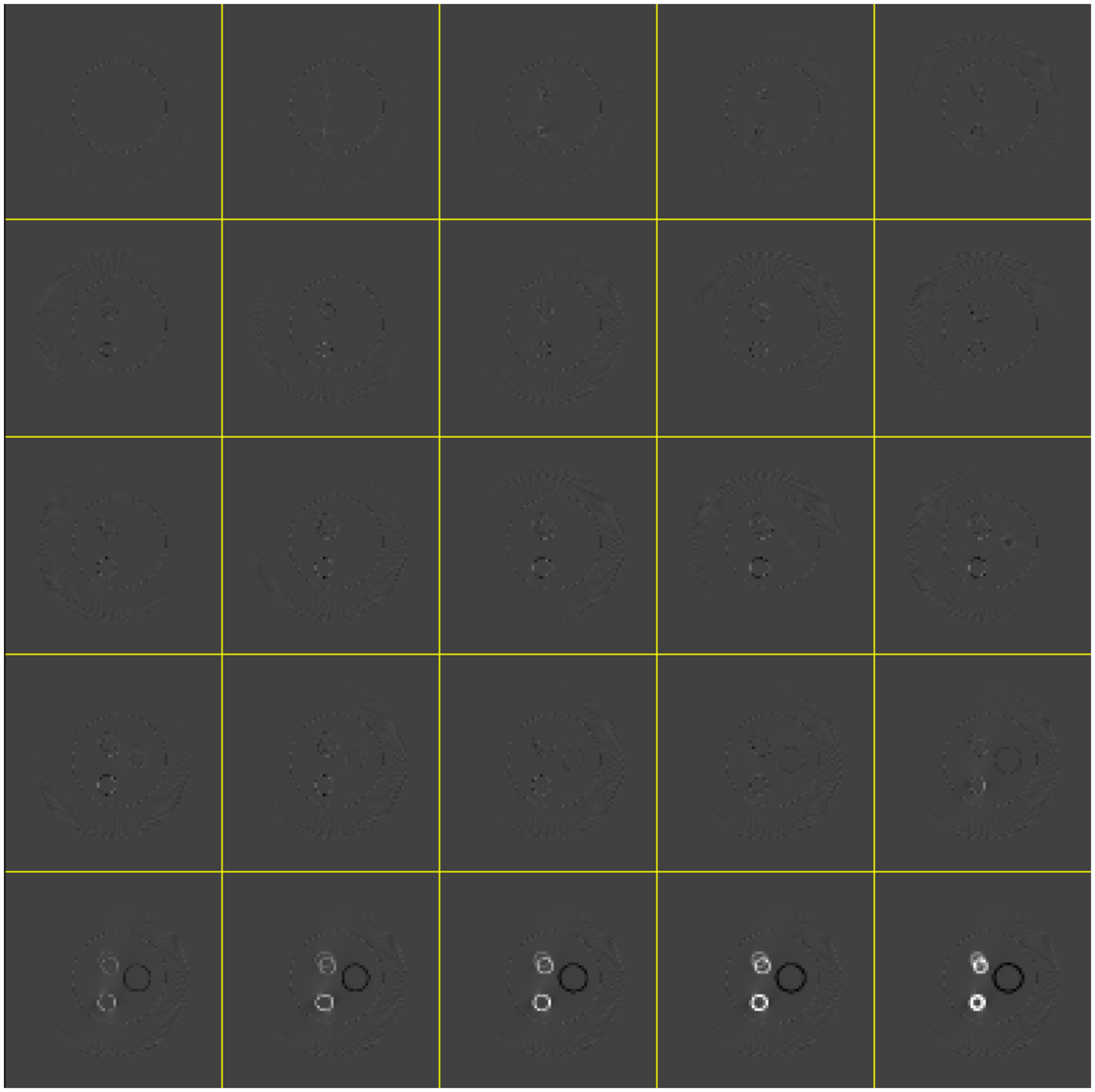}}
  \centerline{(C)}\medskip
\end{minipage}
\caption{Helical scan tested on a simple simulated phantom. Pitch factor is 0.5 in this phantom data. (A) The original phantom.
(B) Image reconstructed with the proposed method. (C) Difference between the true image and the reconstructed image.}
\label{fig:helicaltest}
\end{figure*}

In this section we present results obtained using the proposed algorithm with fan and helical cone beam geometries
using a Shepp-Logan software phantom available in MATLAB (MathWorks, Massachusetts, USA),
a custom made phantom which mimics different cardiac plaques, and a chest scan from a hospital patient. \\
MATLAB was used to calculate the equiangular fan beam projections through the  Shepp-Logan phantom.
The X-ray projections of the plaque phantom and the patient were taken using a Toshiba Aquilion
ONE$^{\copyright}$ scanner (Toronto General Hospital, Canada), which has 896 detectors and 320 rows of detectors.
The scanner gathers data from 1200 projections in each $360^{\circ}$ rotation. When images were reconstructed from fewer than
the 1200 projections the projection views were selected equiangularly. For all the scan protocols used the X-ray tube
current-exposure time product was 50mAs and the peak voltage was 120kV. Data from the central row of a volumetric scan
on one single rotation served as the fan beam data. To simplify the EAW calculation, $\alpha$ was chosen to be a diagonal
matrix whose elements were $\alpha_i \propto 1/d_i$ so that the elements of EAW will be $c_i = 1/(1+\epsilon_i)$.\\
Figure \ref{fig:phan3methodcompare} compares the Shepp-Logan phantom reconstructed from 128 projections using 1) the
inverse pseudo polar Fourier transform (using the least squares method), 2) an iterative soft  threshold-based method
(TwIST) \cite{TwIST}, and 3) the proposed FCSA-LEM method. Based on the same phantom Figure \ref{fig:NormalizedError_FBPvsCS}
compares the accuracy of the reconstruction error for all three methods as the number of projections is varied from 50 to 1000. \\
Both of these figures show that using EAW improves the recovery accuracy. In particular  Figure \ref{fig:NormalizedError_FBPvsCS}
shows that the use of more than 256 projections for a 512$\times$512 image does not significantly affect the reconstruction accuracy.\\
A major improvement in the proposed method, compared to the other CS-based reconstruction methods, is its much reduced computational burden.
Figure \ref{fig:TimeFBPvsCSLog} compares the recovery time of 1) filtered back projection (FBP), 2) the proposed method (FCSA-LEM), and 3) an ART-TV based method \cite{PICCS2008,PICCS2012},
which is basically an algebraic reconstruction followed by a TV smoothing at each step. The computer used for the simulation is a desktop i5 computer with 16GB of RAM.
Using this computer we could not use ART-TV methods with a resolution higher than 128$\times$128 pixels due to memory constraints in MATLAB.
It can be seen that the recovery time for the proposed method gets closer to the time of FBP as the image approaches 1024$\times$1024 pixels. \\
Figure \ref{fig:phantest} compares the plaque phantom reconstructed with FBP from 1200 projections and the phantom reconstructed with
the proposed method from 256 projections. It can be seen that the image quality is almost the same with an error less than 1\%.\\
Reconstructions of a chest CT scan from a hospital patient using FBP from 1200 projections and the proposed method from 256 projections is shown in Figure \ref{fig:patienttest}.
It is evident that the image reconstructed with the proposed method has almost the same quality as FBP, which has about 5 times more projections,
\emph{i.e.} 5 times the radiation dose. While the image from the proposed method is a little blurry compared to FBP reconstructed image, but all the details are preserved. \\
Figure \ref{fig:helicaltest} shows a simple simulated phantom reconstructed by the proposed helical reconstruction method.
The helix source position is defined as $\psi = [R\cos(\varphi), R\sin(\varphi), P\frac{\varphi}{2\pi}]$ and in this test pitch factor $P  = 0.5$.
As can be seen, aside from the start or end of the scan the reconstruction is almost perfect.
However, when the image is close to one of the endpoints the error of rebinning increases and as a result the image reconstruction error increases.
%% --------------------------------------------------------------------------------------------
%%                      Conclusion
%% --------------------------------------------------------------------------------------------
\section{Conclusion}\label{sec:conclusion}
It has been shown that CT reconstruction can be statistically modeled as a weighted compressed sensing optimization problem.
The proposed model was derived from a MAP model of CT imaging with sparsity and piecewise linearity constraints.
To solve the proposed model a fast CS recovery method was proposed in which pseudo polar Fourier transform was used as
the measurement function to reduce the computational complexity. Moreover, to be able to reconstruct CT images from fan beam and helical cone beam projections,
rebinning to parallel beams was used. To adapt the proposed CS recovery method to the interpolation error, a weighting approach (EAW) was proposed,
in which the weights accounted for the measurement noise and interpolation errors. This enabled CT images to be reconstructed from a reduced number
of fan or helical cone beam X-ray projections. It was shown that using EAW improves the reconstruction quality substantially. For instance,
a 512$\times$512 Shepp-Logan phantom reconstructed with 128 projections using a conventional CS method had $\sim 10\%$ error. However, using the
same data with our new method the reconstruction error was as low as $\sim 1\%$. The proposed weighted CS-CT reconstruction model was solved with
a proposed FCSA-EM based method, called FCSA-LEM. The low computational complexity of our FCSA-LEM method made fast recovery of the CT images possible.
For example, we were able to recover a 512$\times$512 image in less than 20 sec on a desktop computer without numerical optimizations,
thus our proposed method may be among the first CS-CT methods whose computational complexity is within the realm of what could be clinically relevant today.
%-------------------------------------------------------------------------
\\ \\
\renewcommand{\abstractname}{Acknowledgements}
\begin{abstract}
\textnormal{
We thank the Canadian Mitacs-Accelerate program and Toshiba Canada for
partial financial support. RSCC wishes to thank the Natural Sciences and
Engineering Council of Canada for support under grant \#3247-2012, and SMH
is grateful for the award of a Loo Geok Graduate Scholarship.}
\end{abstract}
% -------------------------------------------------------------------------
\bibliographystyle{IEEEbib}
\bibliography{refsnew}

\end{document}